\documentclass[12pt]{iopart} 
\usepackage{epsfig}
\usepackage{array}
\usepackage{graphicx}
\usepackage{bm}
\usepackage{epstopdf}
\usepackage{amssymb}

\begin{document}
\title{Evolution of 3D Boson Stars with Waveform Extraction}

%----------
% Authors
%----------
\author{Jayashree Balakrishna$^{1}$, Ruxandra Bondarescu$^{2}$, Gregory 
Daues$^{3}$, F. Siddhartha Guzm\'an$^{4}$ and Edward Seidel$^{5,6,7}$}
\address{$^{1}$Harris-Stowe State University, St.\ Louis, MO USA.\\
$^{2}$Cornell University, Ithaca, NY USA.\\
$^{3}$National Center for Supercomputing Applications, Urbana, IL  USA.\\
$^{4}$Instituto de F\'{\i}sica y Matem\'{a}ticas, Universidad Michoacana de San Nicol\'as de Hidalgo. Edificio C-3, Cd. Universitaria. C. P. 58040 Morelia, Michoac\'{a}n, M\'{e}xico.\\
$^{5}$Center for Computation and Technology, 302 Johnston Hall, Louisiana State University, Baton Rouge, LA 70803.\\
$^{6}$Department of Physics and Astronomy, 202 Nicholson Hall, Louisiana State University, Baton Rouge, LA 70803.\\
$^{7}$Max Planck Institut f\"ur Gravitationsphysik, Albert Einstein Institut, Am M\"uhlenberg 1, 14476 Golm, Germany.\\}
\date{\today}
%-----------------

\begin{abstract}

Numerical results from a study of boson stars under nonspherical perturbations using 
a fully general relativistic 3D code 
are presented together with the analysis of emitted gravitational radiation. 
We have constructed a simulation code suitable for the study of scalar fields in 
space-times of general symmetry by bringing together components for addressing the
 initial value problem, the full evolution system and the detection and analysis of 
 gravitational waves. 
Within a series of numerical simulations, we explicitly extract the Zerilli and 
Newman-Penrose scalar $\Psi_4$ gravitational waveforms when the stars are 
subjected to different types of perturbations. Boson star systems have rapidly 
decaying nonradial quasinormal modes and thus the complete gravitational waveform 
could be extracted for all configurations studied. The gravitational waves emitted from  
stable, critical, and unstable boson star configurations are analyzed and the 
numerically observed quasinormal mode frequencies are compared with known linear 
perturbation results.  The superposition of the high frequency nonspherical modes 
on the lower frequency spherical modes was observed in the metric oscillations when 
perturbations with radial and nonradial components were applied. The collapse of 
unstable boson stars to black holes was simulated. The apparent horizons were 
observed to be slightly nonspherical when initially detected and became spherical 
as the system evolved. The application of nonradial perturbations proportional to
spherical harmonics is observed not to affect the collapse time. 
An unstable star subjected to a large perturbation was observed to migrate to a stable configuration. 

\end{abstract}
\maketitle
\section{Introduction}

Spin zero particles play an important role in particle physics and 
cosmological models. Studies of the early universe have suggested that 
these types of bosons may have played significant roles in determining its 
history. These roles range from the inflaton,  a scalar field that could be 
responsible for the inflation of the universe at very early 
stages \cite{inflaton}, up to the quintessence, which is a scalar field 
used to model the dark energy component of the current universe 
\cite{quintessence}. Bosonic particles could come together through some kind of 
a Jeans instability mechanism to form gravitationally bounded objects such 
as boson stars (complex scalar field) or oscillatons (real 
scalar field) \cite{seidel91a}.  These objects would be held together by 
the balance between the 
attractive force of gravity and the dispersive effects of the uncertainty 
principle (the wave character of the scalar field). Boson stars in the 
Newtonian regime have been considered as candidates for 
dark matter in the context of galactic halos \cite{dark-matter}. In the 
strong field regime these stars have been shown to be reliable models 
for the supermassive objects in the center of some galaxies \cite{diego} 
and it has been shown that they may play the role of black hole candidates 
\cite{guzman05}.
%Scalar fields in general and boson stars in particular, have
%traditionally helped in understanding the gravitational
%critical collapse in spherical symmetry \cite{matt1}, and recent studies point to the
%more general cases of axial symmetry \cite{matt2} and no symmetries in 3D \cite{matt3}.
%Single boson stars as potential sources of gravitational radiation are considered in this paper.

If these objects exist, the process of formation could include 
multipolar components of the gravitational field that should transform
into gravitational waves. Due to post-formation perturbations, boson 
stars could generate gravitational radiation.  The head-on collision of two 
boson stars has been investigated previously as a source of 
gravitational waves \cite{jay-phd}.  When two boson stars collide to form a 
black hole very little scalar radiation is emitted and most of the energy is 
lost in gravitational radiation \cite{jay-phd}. Ryan \cite{ryan} has studied 
the scenario of a boson star of mass greater than several $M_{\odot}$ 
coalescing with a particle such as a small black hole. He finds that this 
system can generate a signal detectable by a gravitational wave interferometer 
such as LIGO and future generation gravitational wave detectors. 
Recently, Kesden {\it et al.} \cite{kesden} have studied the infall of a stellar 
mass compact object onto a supermassive boson star and studied an approximate 
gravitational wave signal, which could point to differences between 
supermassive horizonless objects (like those in Ref.\ \cite{diego}) and 
supermassive black holes in the center of galaxies.

The detection of boson stars, either as compact objects or as
bricks of the dark matter component, necessitates 
studying their behavior under general nonspherical perturbations.
Extracting their particular gravitational wave signatures, under 
general perturbations, requires a fully general relativistic 3D code.  
Boson stars also have an important role in the field of numerical relativity 
aside from their relevance to astrophysics. The performance of long-term stable 
3D simulations of systems with general symmetry and construction of a 
full gravitational wave signal is a timely and challenging problem in 
general relativity.  Boson stars are amenable to stable 3D simulation 
 due to their smooth surface and the absence of singularities, offering a 
suitable testbed for the field.  

The properties of spherically symmetric boson stars are well studied in the 
literature \cite{ruffini,kaup,edw1,jaya,gleiser, lee,jetzer,mielke-schunck}. 
Critical solutions and spherically symmetric boson stars brought to the threshold of black hole formation were studied by Hawley and Choptuik in Ref.\ \cite{matt1}. They performed a comparison between radial quasinormal mode frequencies of boson stars calculated via perturbation theory and those of the numerically constructed critical solutions. An investigation of 
the properties of boson stars beyond the spherically symmetric case was 
performed by Yoshida {\it et al.} \cite{futamase}.  The authors calculate the 
quasinormal mode frequencies corresponding to nonradial pulsations of a boson star.
They perform a linear expansion about the spacetime of a spherically symmetric static 
boson star configuration according to the decomposition into tensor spherical harmonics 
developed by Regge and Wheeler. Their results consist of a series of complex 
quasinormal mode frequencies for even parity perturbations for three models
of stable, critical, and unstable boson star configurations. These modes are damped 
in a short time because of large imaginary parts of the frequencies.

The focus of this paper is the nonspherical perturbation problem of spherical boson stars and the gravitational wave signals generated by such systems. 
The simulations shown here are performed using the fully general 3D simulation code presented in \cite{francisco}, which is based on the Cactus Computational Toolkit \cite{cactus}. The existence of the rapidly damping quasinormal modes predicted by Yoshida {\it et al.} is confirmed numerically. Both the Zerilli and the Newman-Penrose $\Psi_4$ gravitational waveforms are
 fully extracted for the first time for boson stars. The generated waveforms are compared to 
modefits constructed with the $\ell=2$ quasinormal mode frequencies calculated 
by Ref.\ \cite{futamase}. After the waveforms
are extracted numerically, the energy loss to gravitational radiation is 
estimated.

In addition to subjecting stars to perturbations proportional to $\ell=2$ spherical harmonics that allowed  the comparison between our waveforms and the frequency spectrum predicted by Ref. \cite{futamase}, we subjected stars to more physical perturbations and studied (1) the collapse of an unstable boson star to a black hole and (2) the migration of a star from the unstable to the stable branch.  The collapse of an unstable branch boson star was accelerated by applying a mixture of nonradial and radial perturbations. The apparent horizon was observed to be measurably nonspherical when the horizon formed before the complete relaxation of the nonradial modes. The migration is initiated by applying a radial perturbation that significantly reduces the mass of the star. A nonspherical perturbation is superimposed on the spherical one and the system is evolved. At early times (before a single radial oscillation is complete) the full gravitational wave signal is extracted. The migration process was further followed to track the continuing low frequency radial oscillations of the metric function $g_{rr}$ (lasting hundreds of oscillations of  the underlying complex scalar field -- a very long time scale  simulation).  

The paper proceeds by providing a brief overview of the mathematical 
background that describes the evolution system and the formulation of the initial value problem. 
Sec.~\ref{methods} discusses the different types of perturbation methods and the different boson star models considered. The results of simulations are then presented in Sec.\ \ref{pert} for stable, critical and unstable boson star configurations under small nonradial perturbations. 
The collapse of an unstable star to a black hole and the formation of the apparent horizon are subsequently presented in
Sec.\ \ref{blackhole}. Finally, the migration of an unstable branch boson star to the stable branch under explicit radial 
and nonradial perturbations is detailed in Sec.\ \ref{migration}. The results are summarized in the conclusion.

\section{Mathematical Background}

\subsection{General Relativistic Evolution System} 
\label{sec:genrel}

The action describing a self-gravitating complex scalar field in a 
curved spacetime is given by

\begin{eqnarray}
I = \int d^4 x \sqrt{-g} \left( \frac{1}{16 \pi}R \, 
    -\frac{1}{2} [ g^{\mu \nu} 
    \partial_{\mu} \Phi^* \, \partial_{\nu} \Phi  
    + V(|\Phi|^2) ]  \right) 
\label{action}
\end{eqnarray}

\noindent where $R$ is the Ricci scalar, $g_{\mu\nu}$ is the metric 
of the spacetime, $g$ is the determinant of the metric, $\Phi$ is the 
scalar field, $V$ its potential of self-interaction, and the geometric 
units $G=c=1$ have been used. The variation of this action with respect to 
the scalar field leads to the Klein-Gordon equation for the complex scalar 
field, which can be written as

\begin{equation}
\Phi^{; \mu} {}_{;\mu} - \frac{dV}{d|\Phi|^2}\Phi = 0 . 
\label{kg-covariant}
\end{equation} 

\noindent When the variation of Eq.\ (1) is made with respect to the metric 
$g^{\mu\nu}$, the Einstein's equations $G_{\mu\nu}= 8\pi T_{\mu\nu}$ 
arise, and the resulting stress energy tensor reads

\begin{equation}
T_{\mu \nu} = \frac{1}{2}[\partial_{\mu} \Phi^{*} \partial_{\nu}\Phi +
\partial_{\mu} \Phi \partial_{\nu}\Phi^{*}] -\frac{1}{2}g_{\mu \nu}
[\Phi^{*,\eta} \Phi_{,\eta} + V(|\Phi|^2))].
\label{setensor}
\end{equation}

\noindent In the present manuscript we focus on the free-field case, for 
which the potential is $V=m^2|\Phi|^2$, where $m$ is interpreted as 
the mass of the field.

In order to find solutions to the Einstein-Klein-Gordon 
system of equations we use the 3+1 decomposition of Einstein's equations, 
for which the line element can be written as

\begin{equation}
   d s^{2}  = - \alpha^2  dt^2 + \gamma_{ij}  (dx^i + \beta^{i} dt) (dx^j 
              + \beta^{j} dt)  
\label{lineelem}
\end{equation}

\noindent
where $\gamma_{ij}$ is the 3-dimensional metric; from now on latin 
indices label the three spatial coordinates. The functions $ \alpha $ 
and $ \beta^{i}$ in Eq. (\ref{lineelem}) are freely specifiable gauge 
parameters, known as the lapse function and the shift vector respectively. 
The determinant of the 3-metric $\gamma$ is defined as $\gamma = 
\gamma_{ij} \gamma^{ij}$. Throughout this paper the standard general 
relativity notation is used. The Greek indices run from 0 to 3 and the 
Latin indices run from 1 to 3. 

The Klein-Gordon equation can be written as a first-order evolution system 
by first splitting the scalar field into its real and imaginary parts: 
$\Phi = \phi_1 + i\phi_2$, and then defining eight new variables in terms 
of combinations of their derivatives:  $\Pi = \pi_1 + i \pi_2$ and $ 
\psi_a = \psi_{1a} + i \psi_{2a}$ with $ \pi_1= (\sqrt{\gamma}/\alpha) 
(\partial_t \phi_1 - \beta^c \partial_c \phi_1) $ and  
$\psi_{1a}=\partial_a \phi_1$ and similarly  $(1\to2)$. With this 
notation the evolution equations become 

\begin{eqnarray}
 \partial_t \phi_1 &=&  \frac{\alpha}{\gamma^{\frac{1}{2}}} \pi_1 + 
\beta^j \psi_{1j} \\\nonumber
 \partial_t \psi_{1a} &=& \partial_a( \frac{\alpha}{\gamma^{\frac{1}{2}}} 
\pi_1 + \beta^j \psi_{1j}) \\\nonumber
 \partial_t \pi_1 &=& \partial_j (\alpha \sqrt{\gamma} \phi_1^j) 
  - \frac{1}{2} \alpha  \sqrt{\gamma} \frac{\partial V}{\partial 
  \vert \Phi \vert^2} \phi_1 \label{kg-equations}
\end{eqnarray} 

\noindent
and $(1\to2)$. On the other hand, the geometry of the spacetime is evolved 
using the BSSN formulation of the 3+1 decomposition. According to this 
formulation, the variables to be evolved are 
$\Psi = \ln(\gamma_{ij} \gamma^{ij})/12$, 
$\tilde{\gamma}_{ij} = e^{-4\Psi}\gamma_{ij}$, $K = \gamma^{ij}K_{ij}$, 
$\tilde{A}_{ij}=e^{-4\Psi}(K_{ij}-\gamma_{ij} K/3)$ and 
the contracted Christoffel symbols 
$\tilde{\Gamma}^{i}=\tilde{\gamma}^{jk}\Gamma^{i}_{jk}$,
instead of the usual ADM variables $\gamma_{ij}$ and $K_{ij}$. The 
evolution equations of these new variables are described in Refs. \cite{bssn, 
stu}:

\begin{eqnarray}
\partial_t \Psi &=& - \frac{1}{6} \alpha K \label{BSSN-MoL/eq:evolphi}\\
\partial_t \tilde{\gamma}_{ij} &=& - 2 \alpha \tilde{A}_{ij}
\label{BSSN-MoL/eq:evolg} \\
\partial_t K &=& - \gamma^{ij} D_i D_j \alpha  + \alpha \left[
        \tilde{A}_{ij} \tilde{A}^{ij} + \frac{1}{3} K^2 + \frac{1}{2}
        \left( -T^{t}{}_{t} + T \right) \right]
\label{BSSN-MoL/eq:evolK}\\
\partial_t \tilde{A}_{ij} &=& e^{-4 \Psi} \left[
 - D_i D_j \alpha + \alpha \left( R_{ij} - T_{ij} \right) \right]^{TF}
\noindent\\
        && + \alpha \left( K \tilde{A}_{ij} - 2 \tilde{A}_{il}
\tilde{A}_j^l
        \right) \label{BSSN-MoL/eq:evolA}\\
\frac{\partial}{\partial t} \tilde \Gamma^i
&=& - 2 \tilde A^{ij} \alpha_{,j} + 2 \alpha \Big(
\tilde \Gamma^i_{jk} \tilde A^{kj}                              \nonumber \\
&& - \frac{2}{3} \tilde \gamma^{ij} K_{,j}
- \tilde \gamma^{ij} T_{j t} + 6 \tilde A^{ij} \phi_{,j} \Big)
                                                                \nonumber \\
&& - \frac{\partial}{\partial x^j} \Big(
\beta^l \tilde \gamma^{ij}_{~~,l}
- 2 \tilde \gamma^{m(j} \beta^{i)}_{~,m}
+ \frac{2}{3} \tilde \gamma^{ij} \beta^l_{~,l} \Big) .
\label{BSSN-MoL/eq:evolGamma2}
\end{eqnarray}

\noindent where $D_i$ is the covariant derivative in the spatial
hypersurface, $T$ is the trace of the stress-energy 
tensor~(\ref{setensor}) and the label $TF$ indicates the trace-free part 
of the quantity in brackets.
The coupling between the evolution of the BSSN variables and 
the variables describing the evolution of the scalar field is first order. 
That is, Eqs.\ (5) are solved using the method of lines with a 
modified version of the second order iterative Crank-Nicholson (ICN) 
integrator (see Ref.\ \cite{urena2}). After a full time step the stress-energy 
tensor in Eq.\ (3) is calculated and used to solve the BSSN evolution equations 
with an independent evolution loop based on the standard second order ICN 
\cite{icn}.

In addition to the formulation of Einstein's equations and the first order 
form of the Klein-Gordon equation we use certain gauge choices to 
determine the lapse function of Eq.\ (4) (the shift vector is zero in 
all of our simulations). The evolutions presented in Sec.\ \ref{pert} were 
carried out using the $1+\log$ slicing condition with the lapse given by $\partial_t 
\alpha=-2\alpha K$, where $K$ is the trace of the extrinsic curvature. 
The more dynamic evolutions in Secs.\ \ref{blackhole}  and 
\ref{migration}, black hole formation from boson star collapse and 
migration of an unstable boson star to the stable branch, used a 
combination of both maximal and $1+\log$ slicing. Maximal slicing was 
enforced using a K-driver~\cite{coodslicebala}.  

In order to set up the correct scaled quantities to be evolved we use dimensionless 
variables. For an equilibrium configuration the complex scalar field has a time 
dependence of the form $\Phi(r,t)=\phi(r) e^{i \omega_0 t}$. This implies that the 
stress energy components given in Eq.\ (3) and therefore the geometry of the spacetime 
have to be time-independent, whereas the field is oscillating with the characteristic 
frequency $\omega_0$ that depends on the central 
value of the field $\phi(0)$. This characteristic frequency $\omega_0$ together with the mass of the 
boson $m$ provide the natural dimensionless units, 

\begin{eqnarray}
r_{\rm{code}} &=&m r/M_{\rm{Pl}}^2, \qquad t_{\rm{code}} = \omega_0 t, 
  \\\nonumber
\phi_{\rm{code}} &=&\frac{\sqrt{4 \pi}}{M_{\rm{Pl}}} \phi,\quad 
   \alpha_{\rm{code}} = \alpha \frac{m}{M_{Pl}^2 \omega_0},
\end{eqnarray} 

\noindent which are the ones used within the present analysis. For 
further code details refer to Ref.\ \cite{francisco}.
%For this study the code used is written as a Cactus \cite{cactus} thorn. 

\subsection{Formulation of the Initial Value Problem (IVP)}
\label{ivp}

Performing numerical simulations in the 3+1 ADM formulation requires the 
construction of valid initial data on some constant time hypersurface.   
This task entails finding a numerical solution to the four nonlinear coupled 
elliptic equations given by the Hamiltonian and momentum constraint 
equations of the ADM formalism. Finding solutions for physically interesting 
systems that are not highly symmetric can be very difficult and 
computationally expensive. The presence of matter terms as a source
increases the complexity even further. 

We consider the case of time symmetric initial data where the extrinsic 
curvature $K_{ij} =0$ on the initial hypersurface ($t=0$) and a background 
scalar field configuration $ \phi(r) $ is assumed. Under the condition of 
time symmetry the equations decouple and it becomes sufficient to solve 
only one nonlinear elliptic equation, i.e., the Hamiltonian constraint. 
For a massive complex scalar field the constraint equation takes the form
\begin{equation}
R = 16 \pi \rho(\gamma_{ij}, \phi, \Pi)
\label{ham}
\end{equation}
where the scalar curvature is $ R=\gamma^{ij}R_{ij}$ and $\rho$ 
is the energy density.
Using the York formalism, we introduce a conformal factor $ \Psi $ 
that relates the actual 3-metric $\gamma_{ij}$ to an initial 
guess  ${\tilde \gamma}_{ij}$ through the relation 
$ \gamma_{ij} = \Psi^4 {\tilde \gamma}_{ij} $. 
Under this transformation the scalar curvature is given by
\begin{equation}
R = \frac{1}{\Psi^4} \tilde R - \frac{8}{\Psi^5} {\tilde \Delta} \Psi 
\end{equation}
In the standard York procedure, all the terms within the energy 
density are transformed uniformly using the relation 
$ \rho = \Psi^{-8}  \tilde \rho $. 
This transformation would require finding a new bosonic field 
configuration corresponding to the new energy density $\rho$, 
which is a nontrivial problem. We circumvented this difficulty by 
explicitly writing the dependence of the energy density on the metric 
functions. The conformal transformation is applied term by term 
within $\rho$, leaving the scalar field configuration unchanged.
On the initial hypersurface the Hamiltonian constraint (Eq. (\ref{ham}))
can be rewritten as the nonlinear elliptic equation 
\begin{eqnarray}
 \frac{1}{\Psi^4} \, {\tilde R} - \frac{8}{\Psi^5} \, \tilde{\Delta} \Psi = &16 \pi 
\left[ \frac{1}{2} \frac{1}{\Psi^{12} \tilde \gamma} \, \Pi_1^2 
- \frac{1}{2} \frac{1}{\Psi^4} \, {\tilde \gamma}^{ij}  \phi_{1,i} \, \phi_{1,j} 
\right. \nonumber \\
&\left. - \frac{1}{2} m^2 \phi_1^2 + (\phi_1 \mapsto \phi_2) \right]
\label{constraint}
\end{eqnarray}
We start with an initial guess and then solve the constraint equation for the 
conformal factor $\Psi$. This formulation of the IVP problem was first used in Balakrishna's Ph.D. 
thesis for a variety of physical situations related to boson stars~\cite{jay-phd}. %The 

\section{Evolution of Perturbed Boson Stars}
\label{main}
\begin{figure}[ht]
\begin{center}
\leavevmode
\epsfxsize=200pt
\epsfbox{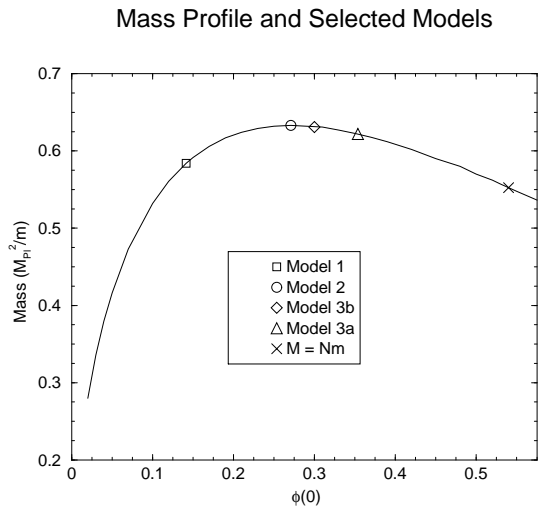}
\caption{
The mass versus central field density $\phi(0)$ for ground state boson 
star configurations is shown. The mass increases to a maximum of $0.633 
M_{\rm{Pl}}^2/m$. This is the critical boson star configuration that separates 
the stable and unstable branches. The points shown in the curve 
correspond to the particular equilibrium configurations analyzed below.}
\renewcommand{\arraystretch}{0.75}
\label{mass_profile}
\end{center}
\end{figure}

%In the present approach we perturb the initial profile of the wave 
% fucntion corresponding to an equilibrium configuration, for which we follow 
% a generalization of the analysis for spherical perturbations in 
% \cite{jaya}, although there are alternatives involving perturbations with 
% external fields that evolve independently of the scalar field of the star 
% \cite{matt1}.
Table \ref{models} details the four different boson star configurations studied in this paper.
The mass profile plot in Fig.\ \ref{mass_profile} shows the mass
versus central field density $\phi(0)$. All configurations listed in Table \ref{models}  
have negative binding energy ($M < N\,m$). Model 1 is a star on the stable branch, Model 
2 is a critical boson star, and Models 3a and 3b both lie on the unstable branch.
These four models are perturbed in different ways, evolved, and analyzed as
described below.

\begin{table}[h]
\begin{center}
\begin{tabular}{|>{\footnotesize}p{0.4in}>{\footnotesize}p{0.4in}>{\footnotesize}p{0.3in}>{\footnotesize}p{0.3in}>{\footnotesize}p{0.7in}|}  \hline
\multicolumn{5}{|c|}{Boson Star Configurations} \\ \hline
Model & $\phi(0)$ &  $M$&$\omega_0$ &$R_{95}$   \\ \hline \hline
 1    & 0.1414  & 0.584  &0.91 & 9.10         \\
 2   & 0.271     & 0.633   &0.85 & 6.03          \\
 3a & 0.3536   & 0.622   &0.82 & 5.05          \\
 3b & 0.300      & 0.631  & 0.84 & 5.65           \\  \hline
\end{tabular}
\end{center}
\caption{Physical characteristics of the boson star configurations studied in this paper are listed. The field $\phi$ is in units of
 $(1/\sqrt{4 \pi}) M_{\rm{Pl}}$. The unperturbed mass $M$ of the boson star is in units of $M_{\rm{Pl}}^2/m$. The equilibrium scalar field
 frequency $\omega_0$ is in units of $m/M_{\rm{Pl}}^2$ and the radius $R_{95}$ has units of $M_{\rm{Pl}}^2/m$. By the radius of the boson star we are referring to the $95 \% $ mass radius of the star. $M_{\rm{Pl}}$  and $m$ are the Planck mass and the mass of the scalar field, respectively.}
\label{models}
\end{table}

All simulations presented in this paper are performed using octant symmetry.
More general simulations on a full grid are the subject of future work. This will allow the
study of the effect of using octant symmetry on the accuracy and stability of boson star simulations.
The waveforms are calculated using the Cactus Extract module for the Zerilli function and the Cactus Psikadelia module for $\Psi_4$ with a radial tetrad choice \cite{cactus}. This particular wave extraction code has been widely used in the past \cite{zer2, zer3, zer4, zer5,  zer9, zer10, zer12}.

The Newman-Penrose complex scalars $\Psi_0$--$\Psi_4$ \cite{penrose_1, penrose_2} 
represent the ten independent components of the Riemann tensor projected onto a basis formed by a null tetrad.  These scalars have asymptotic behavior $\Psi_n \sim r^{n-5}$. Thus, the scalar $\Psi_4$ dominates in the distant wave zone and approximates an outgoing gravitational wave.
 
The Zerilli function \cite{zerilli} is a gravitational waveform that is generated in the study of 
linear perturbations of the Einstein field equations. When the full system of 
the Einstein equations is expanded about a background Schwarzschild metric with a 
small perturbation function $ h_{\mu \nu} $ (proportional to spherical harmonics $Y_{\ell m}$), 
the resulting linear equations contain a set describing even parity perturbations. 
For the Zerilli function $\psi_Z$, the energy loss to gravitational waves can be calculated using
\begin{equation}
\label{zerilli_energy}
E(t) = \frac{1}{32 \pi} \int dt \left( \frac{\partial \psi_Z}{\partial t }\right)^2.
\end{equation}

%Units: $[1/\sqrt{4 \pi} M_{Pl}]$,   [M_{Pl}^2/m], omega[m]  radius [M_{Pl}^2/m] 

\subsection{Methods of Perturbation}
\label{methods}

\begin{figure}[ht]
\begin{center}
\leavevmode
\epsfxsize=200pt
\epsfbox{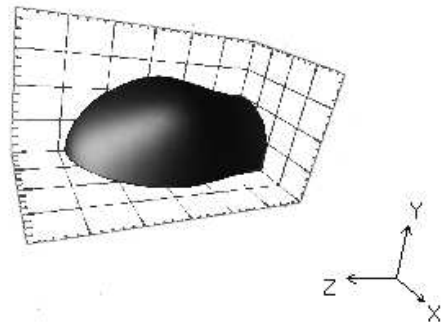}
\caption{Conformal Factor isosurface $\Psi=1.00001$ after the initial value problem solve. 
The perturbation of the field is proportional to the $Y_{20}$ spherical harmonic. The 
dumb-bell shape of the harmonic is apparent in the shape of the conformal factor and 
thus in the initial corrections to the metric.}
\renewcommand{\arraystretch}{0.75}
\label{conformal_factor}
\end{center}
\end{figure}

Various initial data sets that represent different classes of nonspherical perturbations 
of boson star configurations are prepared. The equilibrium field configuration, spherically symmetric solution $\Phi(r)$, is taken as a base and different types of perturbations are added to this configuration. One method involved a scalar field perturbation of the form 
\begin{equation} 
\delta \Phi = \sum_{\ell m} \epsilon_{\ell m} Y_{\ell m}(\theta,\varphi)  f(r) \Phi(r)
\label{ylm_pert}
\end{equation}
where $ f(r) $ is a weighting function. 

Various weighting functions were studied.
\begin{itemize}
\item Uniformly weighted: For this perturbation the weighting function was constant, $f(r)=1$. 
In this case the scalar field was perturbed across the width of the star,  
and was of significant size near the origin where the field density $\phi$ is maximum. 
\item Parabolic weighted : 
\begin{equation}
f(r) =  \left\{ \begin{array}{ll}  \left( \frac{r}{R_p} \right)^2  & r<R_p \\ 1 & r \ge R_p \end{array}  \right.
\end{equation}
(where $R_p$ is a radius at which the perturbation is centered). In this case the perturbation is set
to zero at the origin. This results in a more slight perturbation, away from the central peak of the  field density of the star. 
%\item Gaussian weighted:  A Gaussian radial weighting $f(r) \sim \exp[-(r-r_0)^2/\sigma^2]$ was also investigated. The Gaussian structure mimics realistic, localized physical disturbances. 
\end{itemize} 

More general perturbations involved both radial and nonradial components.
 One such radial-nonradial (R-NR) perturbation type was constructed by giving full weight to 
 a spherically perturbed profile along one direction, while keeping an 
 unperturbed profile in the perpendicular directions and a smooth linear combination of the two  
 in the intermediate region. 
 %Another type involved a numerical perturbation with unequal grid 
%resolution with $\Delta x \ne \Delta y \ne \Delta z$ as well as $\Delta x=\Delta y \ne \Delta z$.
 
  For all the perturbations described in this paper the metric functions are not explicitly perturbed.
  The initial value problem solver starts with the perturbed scalar field configuration described 
above and an initial guess for the metric functions and then solves the constraint Eq.\ (\ref{constraint})
 for the corrected metric functions. The initial
 guess for the metric functions is taken to be the usual spherically symmetric 
 models of Ref. \cite{edw1,ruffini,kaup} obtained by solving a 
 one-dimensional eigenvalue problem for time independent metric functions and for a field 
configuration oscillating with fixed frequency $ \omega_0 $ and constant amplitude at the origin.
These perturbations are similar to those performed by Ref.\ \cite{jaya}. Alternative perturbations
could involve external fields that evolve independently of the scalar field of the star (see 
Ref.\ \cite{matt1}). 

An example of the results obtained from the IVP solver is shown in Fig.\ \ref{conformal_factor}. 
An isosurface of the conformal factor for a perturbation of the field proportional to the $Y_{20}$ 
spherical harmonic is visualized. The prominent dumb-bell shape of the harmonic 
along the $ z $-axis resurfaces in this display of the corrections to the initial metric.

\subsection{Perturbation cases studied}
\label{pert}
\begin{figure}
\begin{center}
\leavevmode
\epsfxsize=200pt
\epsfbox{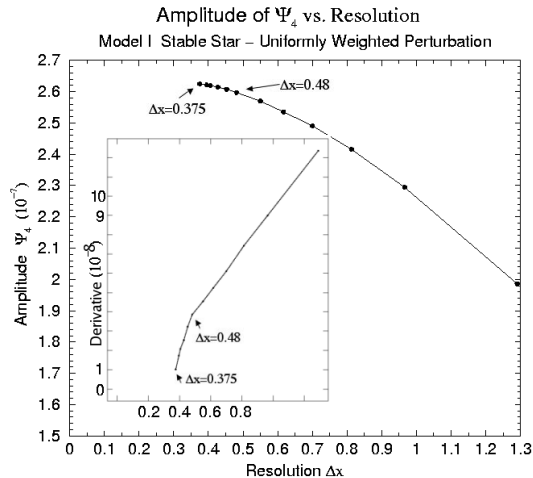}
\caption{The amplitude of the Newman-Penrose scalar $\Psi_4$ is displayed as a function of resolution for a collection of simulations of a Model I stable star. In each simulation the waveform was extracted at a detector located at $r=56.8$. The $\Psi_4$ amplitude of a given simulation is represented by black dot on the plot. The inset shows the first derivative of the amplitude curve. It can be seen that for a resolution better than $\Delta x = \Delta y = \Delta z = 0.48$ the curve flattens and its derivative goes sharply towards zero. A linear extrapolation towards infinite resolution
shows the amplitude at $\Delta x = \Delta y = \Delta z = 0.375$ to vary from the continuum 
limit by only about $1$-$2\%$.}
\renewcommand{\arraystretch}{0.75}
\label{amp}
\end{center}
\end{figure}
% \begin{figure}[ht]
\begin{figure}[t]
\begin{center}
\leavevmode
\epsfxsize=200pt
\epsfbox{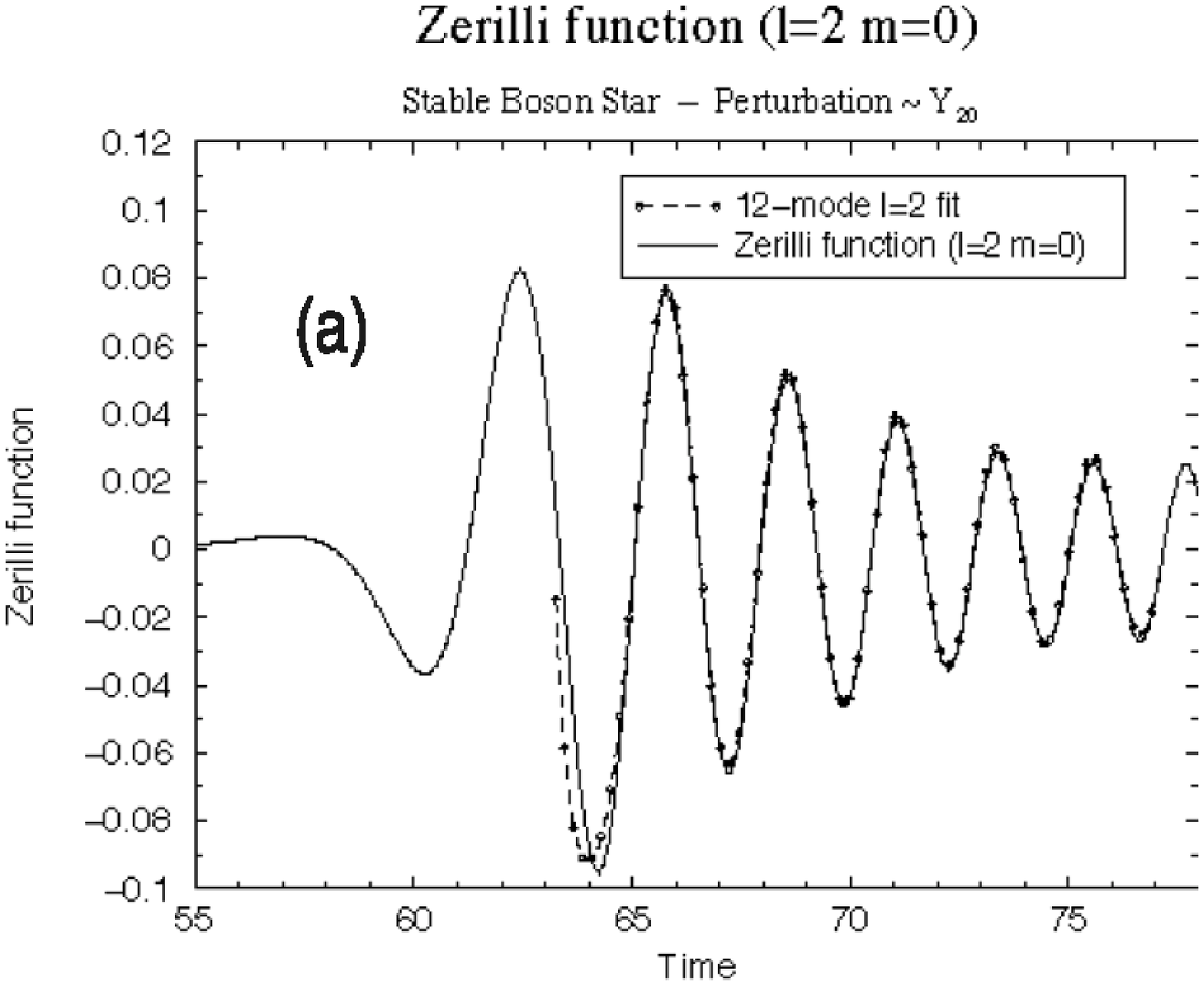}
\epsfxsize=200pt
\epsfbox{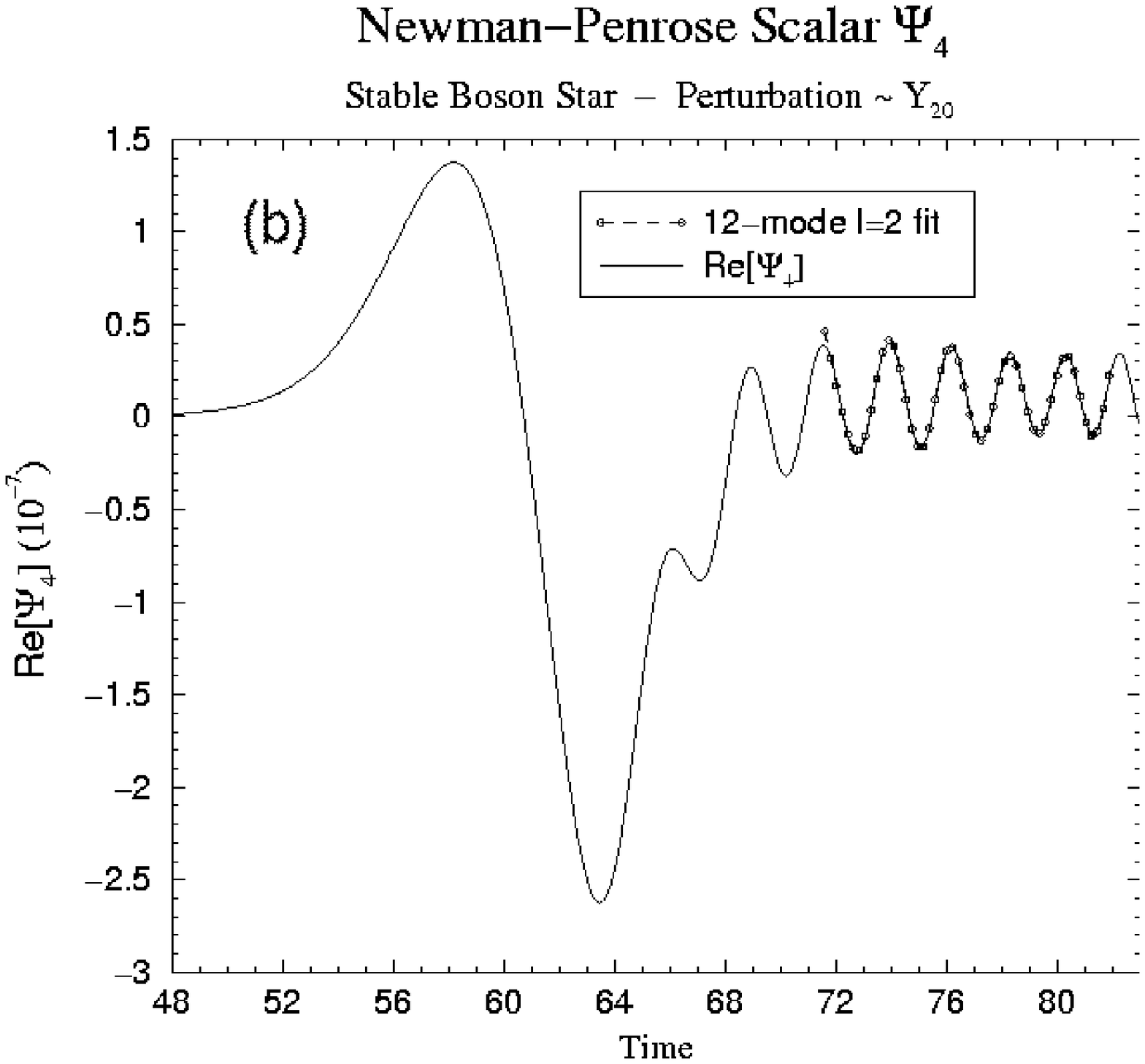}
\epsfxsize=200pt
\epsfbox{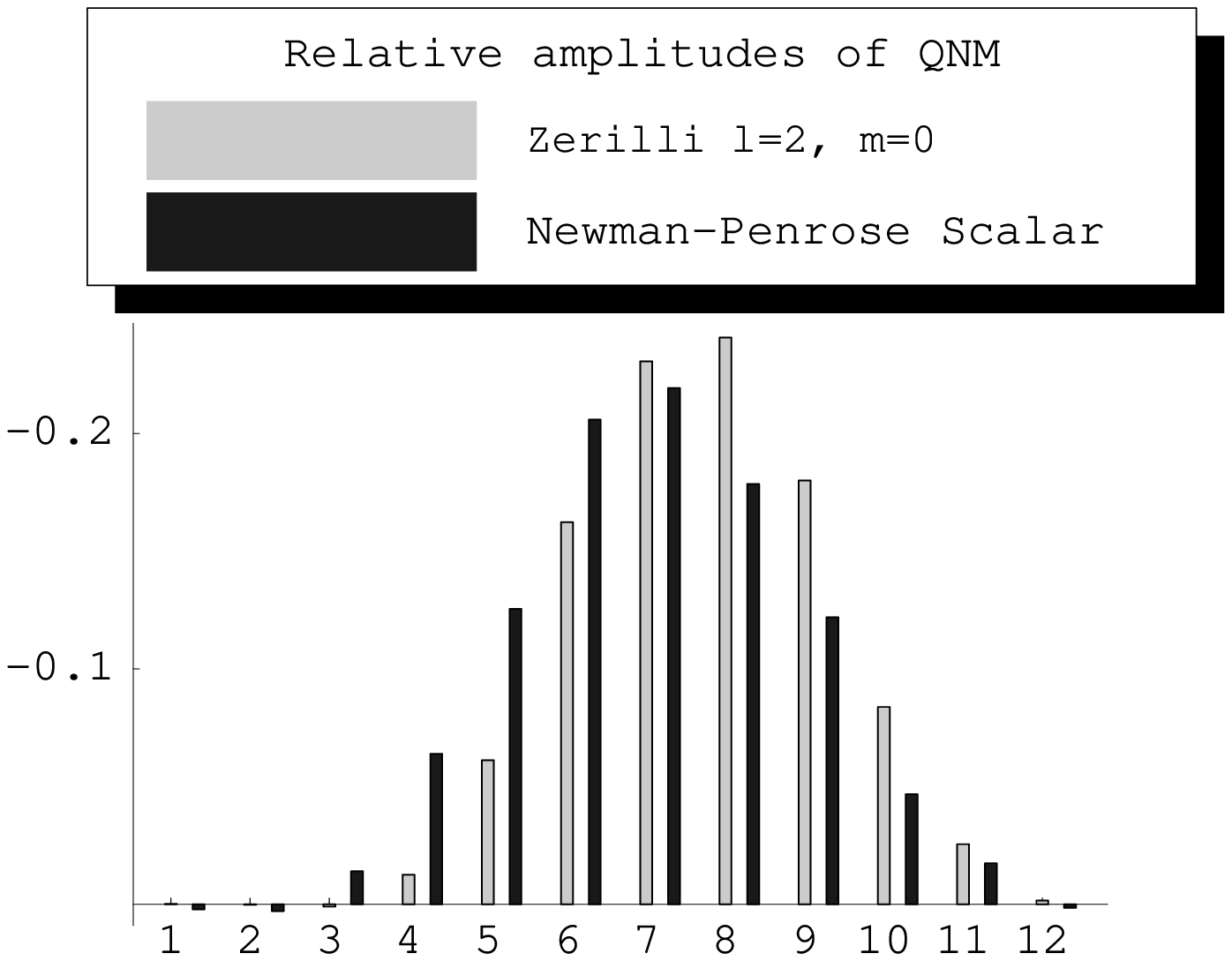}
\caption{A stable boson star (Model 1, Table \ref{models}) is subjected to a small uniformly 
weighted perturbation proportional to $Y_{20}$ with 
$\epsilon_{20}=0.032$. (a) The Zerilli $\ell=2, m=0$ and (b) 
the $\Psi_4$ waveforms are shown in agreement with modefits (for time intervals [$65.0$, $79.2$] and [$71.6$, $81.9$], respectively) using the first twelve quasinormal modes of Ref.\ \cite{futamase}. Both waveforms exhibit a low frequency precursor that likely represents the initial shaking of the star due to an instantaneous finite perturbation. $\Psi_4$ appears to be a more sensitive indicator of the initial dynamics of the star. (c) Coefficients of the modes for the two waveforms are shown to have similar profiles. The contribution of the first two modes is negligible, probably due to the failure of the WKB approximation in determining these modes accurately \cite{futamase}.}
 \renewcommand{\arraystretch}{0.75}
 \renewcommand{\topfraction}{0.6}
\label{zerilli_y20}
\end{center}
\end{figure}

\subsubsection{`Pure' nonspherical perturbations}
\label{nonsph}

The evolution of boson stars under `pure' nonspherical perturbations ($\delta M(r) \sim 0$ 
where $\delta M$ is the mass change) that are proportional 
to spherical harmonics was investigated. Under such perturbations the quasinormal 
modes corresponding to nonradial oscillations were found to have frequencies with both 
real and imaginary parts~\cite{futamase} (even for  configurations that are unstable or 
critical in nature with respect to spherical perturbations). Thus, it was anticipated that 
equilibrium boson star configurations would be stable with respect to nonspherical perturbations. 

A series of simulations of perturbed star configurations was performed by 
taking different values for the coefficients $\epsilon_{\ell m}$ and the weighting function $f(r)$ 
in Eq.\ (\ref{ylm_pert}). These perturbations do not significantly change the mass of the 
star or the number of particles (even for a large nonspherical perturbation 
the mass change is less than $0.01\%$). The gravitational wave content 
is calculated using the even parity quadrupole ($\ell=2$) Zerilli waveforms and the Newman-Penrose scalar $\Psi_4$. The waveforms are computed at several different detector locations in the spacetime.  The detectors had to be located sufficiently far from the center of the 
star for the $\Psi_4$ scalar to dominate the other scalars in the 
Newman-Penrose formalism. Since the boson field tails off exponentially, 
the spacetime in the exterior regions ($r>R_{95}$) becomes increasingly 
close to Schwarzschild as the radius increases. The perturbation formalism 
associated with the Zerilli waveforms is valid when the metric functions describing 
the spacetime are only small perturbations away from those of the Schwarzschild spacetime. Thus detectors measuring Zerilli waveforms should be located at a distance greater than several $R_{95}$ from the center of the star. Simultaneously, the detectors cannot be too close to the boundary in order to avoid the effects of reflection from the edge of the grid. These two requirements coupled with computational limitations on the grid size and the need for good resolution are the main challenges of our numerical simulations. 
 %The perturbations were observed to propagate rapidly away from the star in the form of gravitational radiation. The evolution of the gravitational waves was followed. The star configurations were observed to become spherical after the waves moved out of the computational domain. The profiles of the Zerilli functions and $\Psi_4$ display exponential decay of their 
% amplitudes.

A uniformly weighted perturbation was first considered. 
Two different perturbations were applied to a Model 1 
(Table \ref{models}) stable configuration. One perturbation 
was proportional to the $Y_{20}$ spherical harmonic with constant of proportionality 
$\epsilon_{20}=0.032$. The second perturbation was a linear combination of 
$Y_{20}$ and $Y_{22}$ with coefficients of $\epsilon_{20}=0.032$ and $\epsilon_{22}=0.026$. 
In the beginning several simulations were performed to determine the scale of accurate 
grid resolution. Fig. \ref{amp} shows the amplitude of the real part of the Newman-Penrose $\Psi_4$ scalar for the same grid size at different resolutions 
for the stable star configuration under the perturbation proportional to $Y_{20}$ 
described above. The waveform is extracted at a radial location of $r=56.8$. From 
the figure it can be seen that at coarse resolution there is a strong dependence on 
amplitude. However, for resolutions better than $\Delta x = \Delta y = \Delta z = 0.48$ 
the curve flattens, with the slope heading rapidly towards zero. The amplitude at the 
best resolution $\Delta x= \Delta y = \Delta z = 0.375$ varies from an extrapolation 
to the infinite resolution limit by only about $1$-$2\%$. The final simulation was carried out on a 
$164^3$ grid with a resolution of $\Delta x= \Delta y = \Delta z = 0.375$. The 
L2-norm of the Hamiltonian constraint for this run remained below $1.2 \times10^{-4}$ 
for the duration of the evolution. The L2-norm of the momentum constraints is about an order of magnitude lower than that of the Hamiltonian constraint with a maximum value of $4 \times 10^{-5}$.

Figs.\ \ref{zerilli_y20}(a) and (b) show the $\ell=2, m=0$ Zerilli function 
and the Newman-Penrose scalar $\Psi_4$ from a detector at $r=56.8$. The plots also show a linear least squares regression using the first 
twelve $\ell=2$ quasinormal modes calculated in Ref.\ \cite{futamase}. It is observed that after an initial precursor, the star rings into a linear combination of its quasinormal modes.
The waveforms are observed to be four orders of magnitude above the level of the noise, which is taken to be the signal obtained from a simulation of an unperturbed spherical star configuration with the same grid size and resolution.
Although the perturbation of the field contains only $ \ell=2, m=0 $ spherical harmonics, it is observed that higher order $\ell=4$ Zerilli modes of the gravitational radiation are also triggered. However, these Zerilli modes have sufficiently low amplitude (about an order of magnitude smaller than $\ell=2$ modes) that they do not contribute significantly to the energy loss in gravitational radiation. The waveform $\Psi_4$ contains all nontrivial modes of the star. Nevertheless, the $\ell=2$ modes dominate and we are able to fit the signal with the $\ell=2$ frequencies determined by Ref.~\cite{futamase}. Using Eq.~(\ref{zerilli_energy}) the energy loss in gravitational waves was calculated to be about $0.1 \%$ of the mass of the star. 
%as can be seen from the agreement between the fit and the oscillations of the gravitational waveform. 

In Fig.\ \ref{zerilli_y20}(c) the relative weights of the twelve modes within the Zerilli and $\Psi_4$ 
waveforms for the stable star from Fig.\ \ref{zerilli_y20}(a) and (b) are shown. 
Both waveforms have roughly the same mode content. The slow increase of the imaginary part 
of the modes as the order of the mode increases necessitates the use of numerous modes in our fit. Note that in black hole systems, it is usually sufficient to fit waveforms dominated by quasinormal modes using only the lowest two modes \cite{bhwaveforms}. In this case, however, it is justified to neglect only the modes beyond about twelve, on the grounds that by then the imaginary part is sufficiently large and they damp out quickly. As can be seen in the figure, even the $11^{\rm{th}}$ and $12^{\rm{th}} $ modes hardly contribute.
It is noteworthy that the first two quasinormal modes are almost absent from the fit. 
However, these first modes have been described as inaccurate in Ref.\ \cite{futamase} by 
the authors because of the break down of their WKB approximation in that domain. It should also be noted that values of the eigenfrequencies of Ref.~\cite{futamase} are dependent on their choice 
for a definition of the surface of the boson star. There is no unique definition for this radius,
since the bosonic field extends to infinity with exponential damping. Hence, there is a 
degree of arbitrariness in the position  of the surface and in the exact numerical values 
of the eigenfrequencies.   In their work the surface of the star was artificially placed where 
the value of the field was of order $10^{-5}$. We take more realistic field configurations, which are 
limited by the size of the grid and the resolution of our simulations.

\begin{figure}
\begin{center}
\leavevmode
\epsfxsize=200pt
\epsfbox{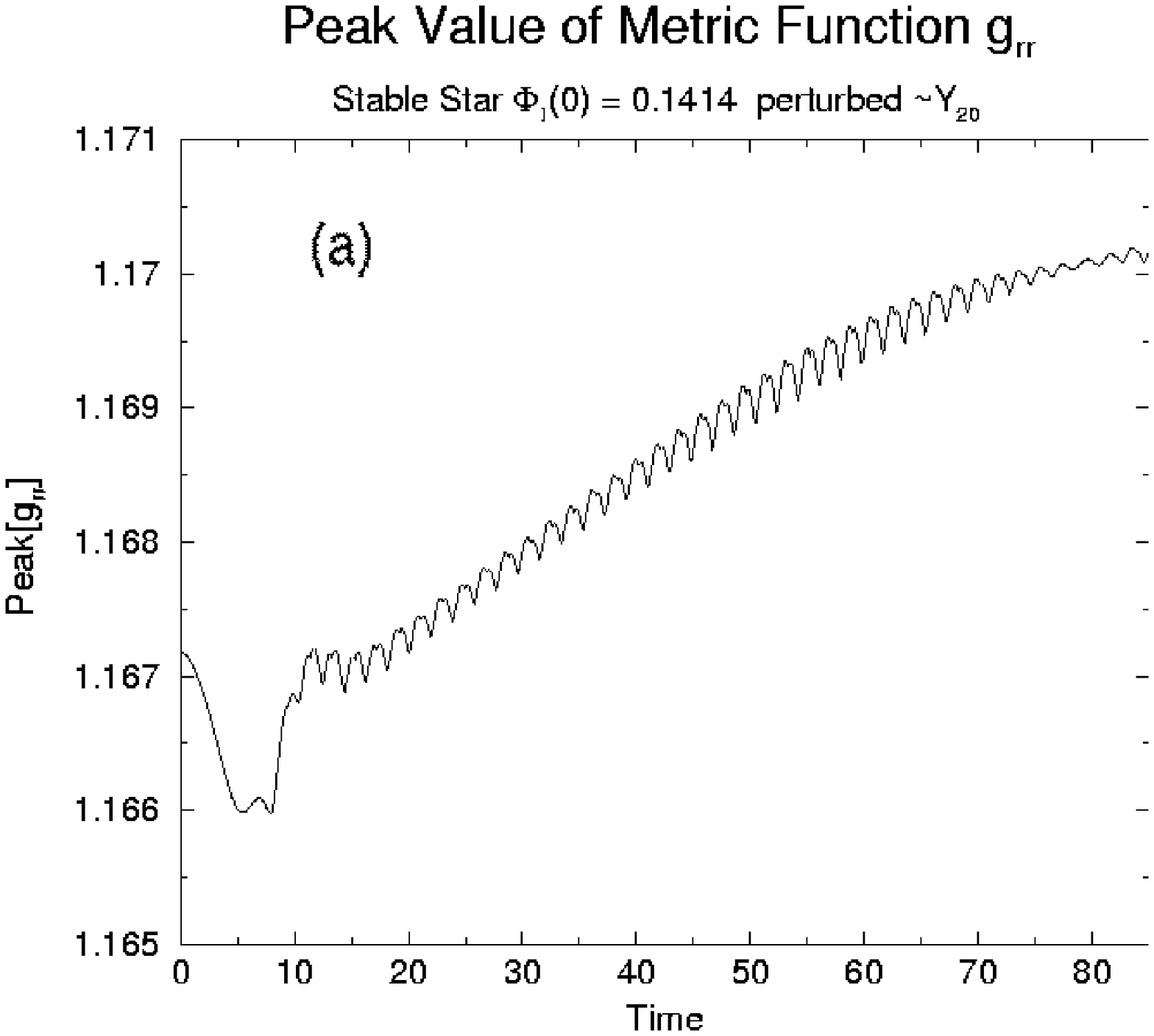}
\epsfxsize=200pt
\epsfbox{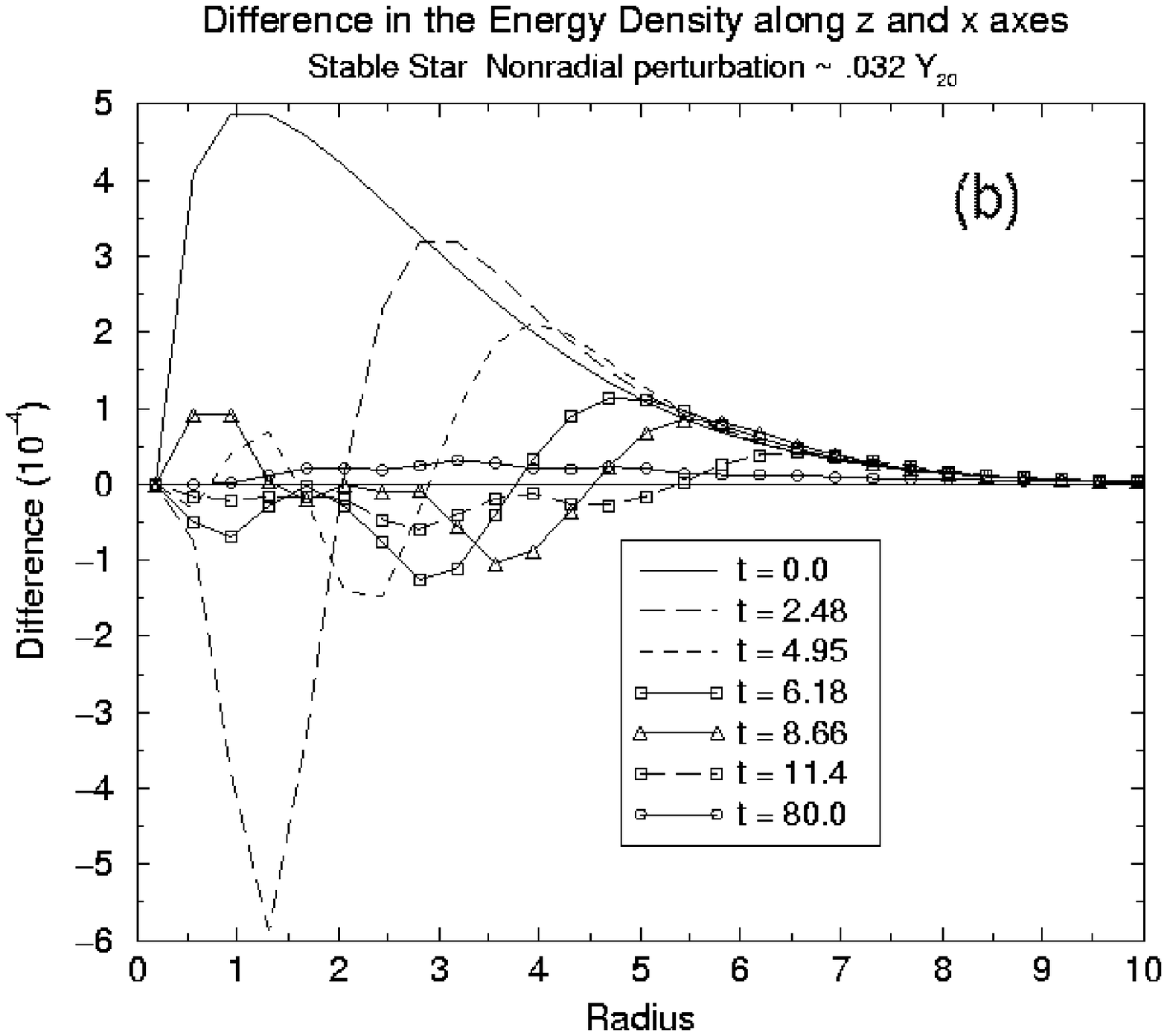}
\caption{ A Model 1 star is perturbed with a uniformly weighted perturbation proportional to $Y_{20}$ with $\epsilon_{20}=0.032$. 
Although no explicit radial perturbation was applied, numerical perturbations due to grid discretization 
(resolution $\Delta x=\Delta y=\Delta z=0.375$) impose a slight radial perturbation on the system. 
The amplitude of this oscillation is seen to converge toward zero as the resolution is improved.
(a) The maximum value of $g_{rr}$ is plotted against time. As a result of the nonspherical perturbation the metric exhibits a high frequency oscillation superimposed on the low frequency one. By $t=80$ the star has undergone only half of its radial oscillation, while the nonradial oscillations have damped off.  
(b) The difference between the density $\rho$ in the $x$ and $z$ directions is shown as a function of time. This difference is an indicator of the degree of asymmetry in the system. It is clearly decreasing as the system evolves and the star becomes spherical after the relaxation of the nonradial modes.
}
\label{gxx}
\end{center}
\end{figure}
%In the continuum limit one should only observe the high frequency oscillations.

The maximum of the radial metric function $g_{rr}$ is shown in Fig.\ \ref{gxx}(a).  The 
nonradial quasinormal mode oscillations are superimposed on the radial 
oscillation. The latter is introduced by grid discretization and its amplitude is resolution
dependent. In the continuum limit one should observe only the high frequency oscillations. Ref.~\cite{francisco} 
has shown that the amplitude of the radial oscillation converges toward zero as the resolution is improved and that in the spherical 
case no high frequency oscillations appear. For this run at a resolution of $\Delta x=\Delta y=\Delta z=0.375$, 
the maximum of the metric has risen only from about $g_{rr \, \rm{max}}=1.166$ to $g_{rr \, \rm{max}}=1.170$. 
The frequency of a radial oscillation of this star configuration was calculated for this simulation to be 
about $0.010$ $m/M_{\rm{Pl}}^2$, which agrees with the result $\sim 0.0097$ $m/M_{\rm{Pl}}^2$ obtained from 
a 1D code for the same configuration.  Since the radial oscillation has a large time period, only 
half of this oscillation has occurred by the end of this run. The damping of the nonradial modes can be 
seen towards the end of the simulation, indicating that the star was settling into a spherical configuration. 
Another simulation that doubled the amplitude of the nonspherical perturbation $\epsilon_{20}$ was performed. 
It was observed that the amplitude of the nonradial oscillations in the metric $g_{rr}$ also doubled. 
This is consistent with the result of  Ref.~\cite{futamase} that the metric perturbation is comparable to 
the scalar field perturbation.

Fig.\ \ref{gxx}(b) shows the difference between the density $\rho$ in the $x$ and $z$ directions as a function of time. This difference is seen to be decreasing by an order of magnitude in a short timescale. Furthermore, the radial location of the peak of the difference function is observed to be moving out as the star becomes more spherical, coinciding with the emission of gravitational radiation.

\begin{figure}[t]
\begin{center}
\leavevmode
\epsfxsize=200pt
\epsfbox{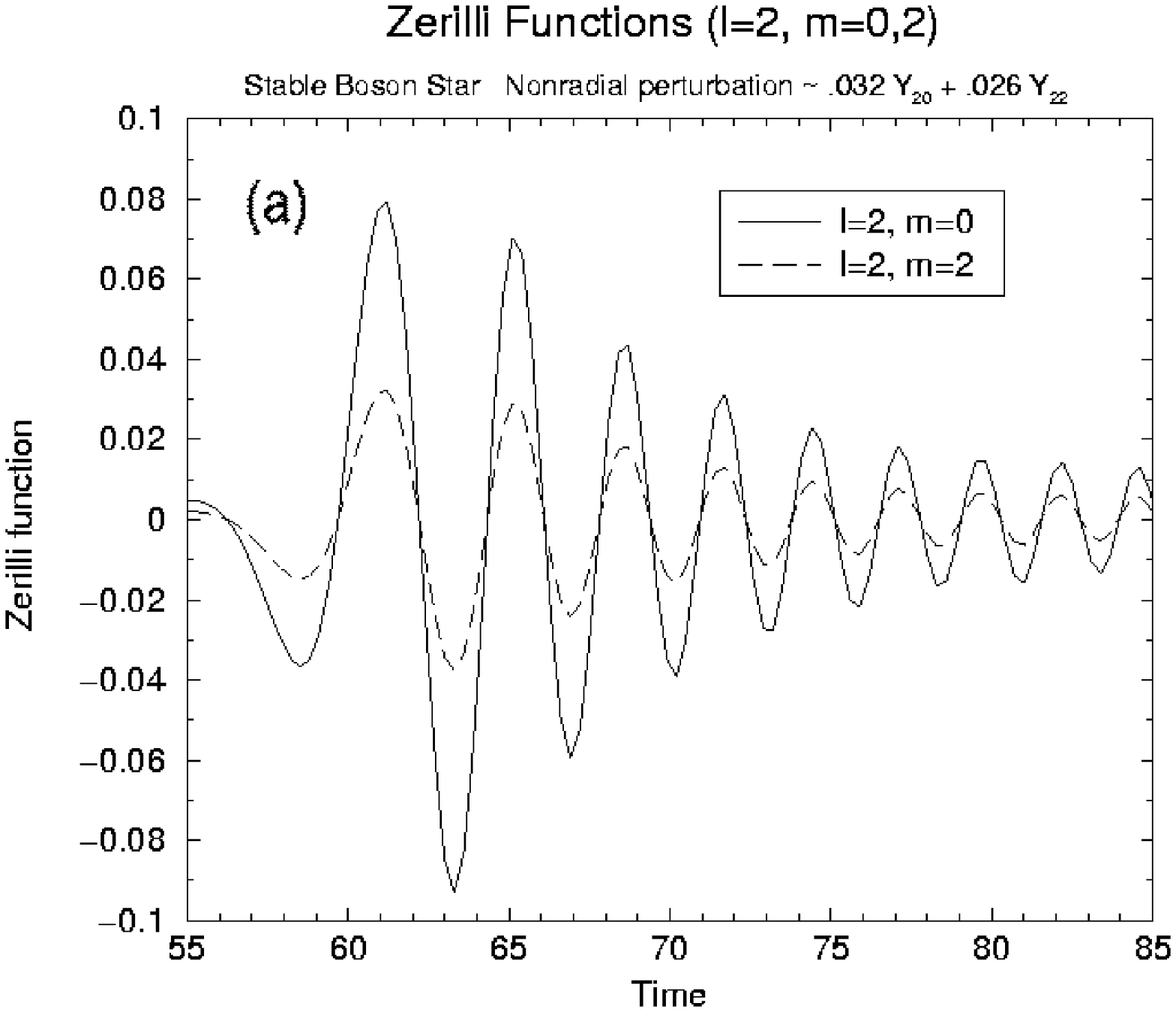}
\epsfxsize=200pt
\epsfbox{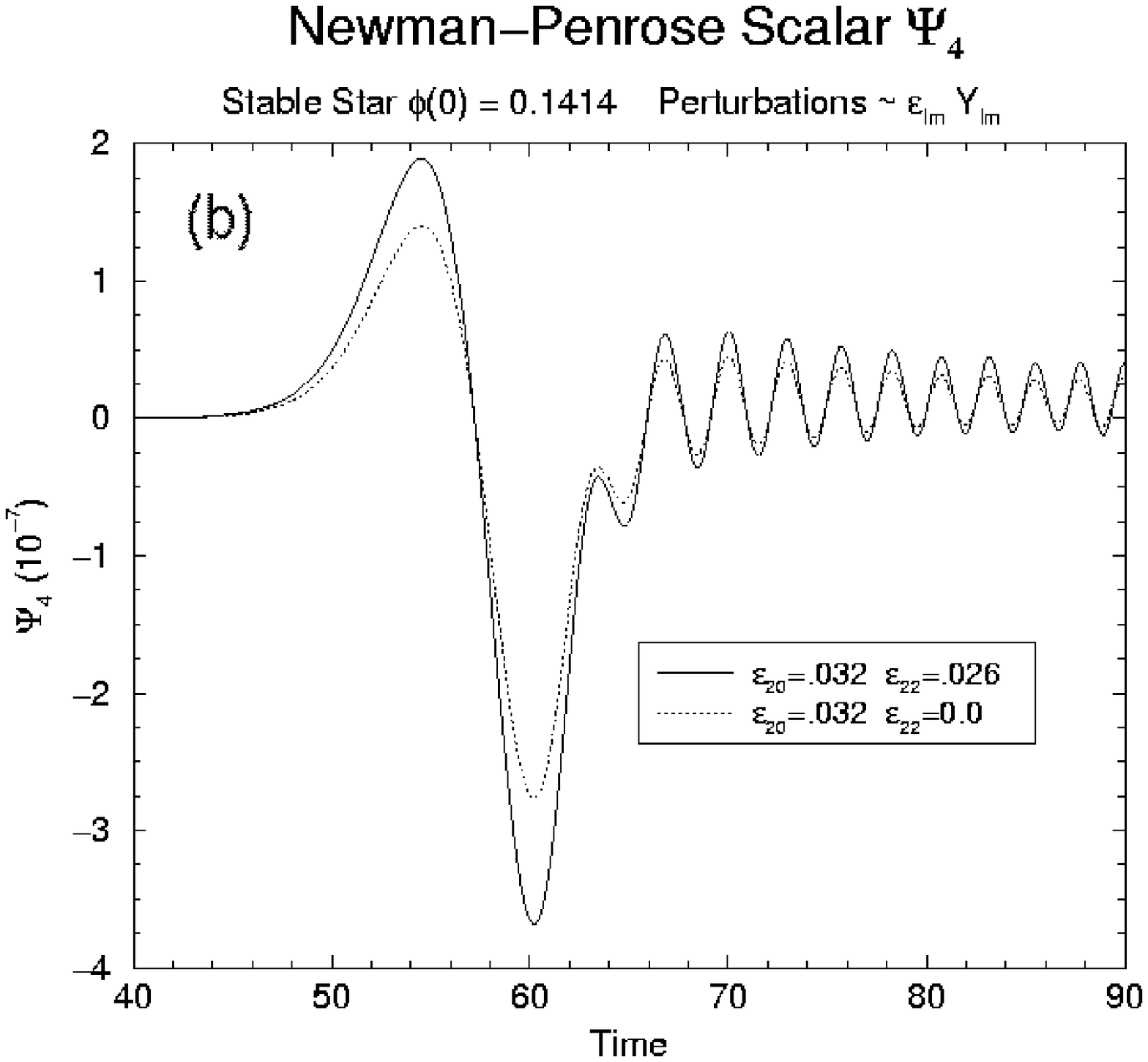}
\caption{(a)  Figure displaying the $\ell =2$ Zerilli waveforms for a Model 1 stable 
star under a small uniformly weighted perturbation proportional to both the $Y_{20}$ and $Y_{22}$ 
spherical harmonics with amplitudes $\epsilon_{20}=0.032$ and $\epsilon_{22}=0.026$. 
The frequency of the Zerilli $\ell=2, m=2$ is identical to the frequency of the $\ell=2, m=0$ Zerilli. 
(b) The $\Psi_4$ waveform for the same star for the perturbation proportional to a linear combination 
of $Y_{20}$ and $Y_{22}$ is compared to the waveform for a $Y_{20}$ perturbation with 
$\epsilon_{20}=0.032$. The very  similar frequencies 
indicate that this configuration has specific nonradial quasinormal mode signatures. 
It can be seen that the $\Psi_4$ signal for the mixed perturbation is larger because 
the Newman-Penrose scalar incorporates more modes.}
\renewcommand{\arraystretch}{0.75}
\label{zerilli_y22}
\end{center}
\end{figure}

 \begin{figure}[t]
 \begin{center}
 \leavevmode
\epsfxsize=200pt
\epsfbox{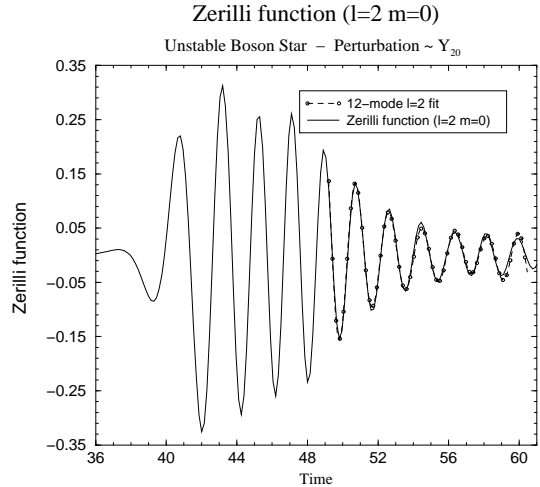}
\caption{The Zerilli waveform for an unstable boson star (Model 3a, Table \ref{models}) under a small uniformly weighted perturbation proportional to the $Y_{20}$ spherical harmonic ($\epsilon_{20}=0.032$) is displayed. The waveform is shown along with modefit applied over the interval [$49.2$, $60.5$].
}
\renewcommand{\arraystretch}{0.75}
\label{crit_unst}
\end{center}
\end{figure}

We now apply a different perturbation to Model 1 and study its evolution. 
In this case a uniformly weighted perturbation proportional to a linear combination 
of $Y_{20}$ and $Y_{22}$ (amplitudes $\epsilon_{20}=0.032$ and $\epsilon_{22}=0.026$, 
respectively) was applied. Fig.~\ref{zerilli_y22}(a) presents the extracted 
Zerilli $\ell=2$ (both $m=0$ and $m=2$) waveforms for a stable star 
(Model 1, Table~\ref{models}). These 
waveforms are observed to have the same frequency. Furthermore, the Zerilli $\ell=2,
m=0$ signal for the star under this mixed $Y_{20}$ and $Y_{22}$ perturbation has the 
same size and frequency as the $\ell=2, m=0$ Zerilli function of the same star under a 
perturbation proportional only to $Y_{20}$ (in both cases $\epsilon_{20}=0.032$).  
This result shows that the addition of other harmonics to the perturbation does not alter 
the original signal and demonstrates the accuracy of the code. Fig.~\ref{zerilli_y22}(b) 
shows a comparison between the Newman-Penrose scalars $\Psi_4$ for the pure 
$Y_{20}$ perturbation and the mixed $Y_{20}$--$Y_{22}$ perturbation. The signals have
the same frequencies, but the signal for the mixed perturbation is larger in size because 
it incorporates more modes.
 
Perturbations proportional to the $Y_{20}$ spherical harmonic with amplitude 
$\epsilon_{20}=0.032$ were studied in the case 
of a critical boson star (Model 2, Table \ref{models}) and an unstable boson star 
(Model~3a, Table~\ref{models}). Since these stars are more compact than the stable star, 
they need better resolution. However, they have a smaller radius and do not need the same 
grid size. A $164^3$ grid with a resolution of $\Delta x=\Delta y=\Delta z=0.25$ was used for both these simulations. For the time interval relevant to our study (that prior to and
encompassing waveform extraction) the simulations proceed accurately with the L2-norm of the Hamiltonian constraint remaining below $4.6 \times 10^{-4}$ for the critical case and below $6.1 \times 10^{-4}$ for the unstable case, respectively.  At late times both stars collapse to black holes due to radial perturbations introduced by grid discretization errors. More detailed simulations of unstable star collapse with apparent horizon analysis are presented in Sec.~\ref{blackhole}.
Fig.\ \ref{crit_unst} displays the $\ell=2, m=0$ Zerilli waveform and modefit for this unstable star configuration with a detector located at $r=35$.  After a fairly large precursor, the star settles into a linear combination of its quasinormal modes with late time agreement between the modefit and numerical waveforms. The extended precursors are representative of complex dynamics within unstable systems.  %Using Eq.~(\ref{zerilli_energy}) the energy loss in gravitational waves was calculated to be about $1 \%$ of the mass of the star for a critical star under the same perturbation and $2.5 \%$ for the unstable star.
%The collapse time seems to be affected only by the resolution of the grid and not by the amplitude of the nonradial perturbation.  More detailed simulations of unstable star collapse with apparent horizon analysis are presented in Sec.~\ref{blackhole}. In that section radial perturbations are explicitly applied, and the sizes of the radial and nonradial perturbations are varied. 
%Another type of perturbation (described in Sec.\ \ref{methods}) for which waveforms were extracted and modefits performed was a small Gaussian weighted perturbation. One motivation for studying a variety of perturbations is to observe 
%whether the system will be dominated by different frequencies from its spectrum 
%of quasinormal modes. The modefit was good, but there was no significant shift in the
%weights of the modes suggesting a characteristic quasinormal mode pattern for a 
%given configuration.

\subsubsection{Perturbation weighted along a chosen axis}
\label{sub1}

\begin{figure}[t]
\begin{center}
\leavevmode
\epsfxsize=200pt
\epsfbox{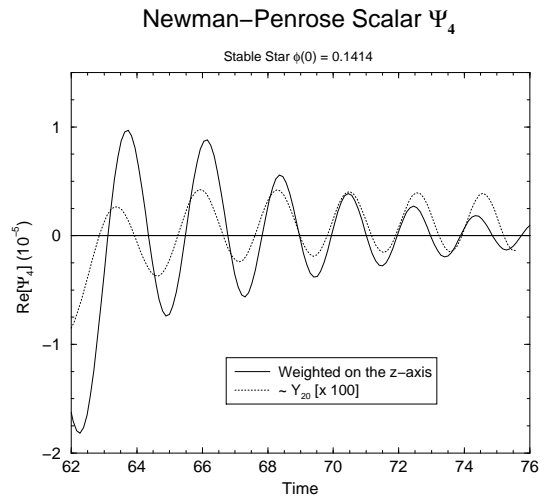}
\caption{Newman-Penrose scalar $\Psi_4$ for a Model 1 star under a R-NR perturbation 
is superimposed on the signal due to the $Y_{20}$ perturbation shown in Fig.~\ref{zerilli_y20}(b).
The amplitude of the latter has been magnified by a factor of $100$ to facilitate the 
comparison. The R-NR perturbation produces a much larger signal. The star did not ring 
in its quasinormal modes initially, but appears to move into these modes after a few 
oscillations. The rapid damping of the gravitational wave signal characteristic to boson stars can be clearly seen.}
\renewcommand{\arraystretch}{0.75}
\label{psi4_old}
\end{center}
\end{figure}

Perturbations with radial and nonradial components as described in Sec.\ \ref{methods} were 
also studied.  In Ref.\ \cite{jay-phd} it was shown that for this type of perturbation the 
dominant radiation is in the form of scalar radiation. The scalar radiation emission process, denoted gravitational cooling \cite{gcooling}, relaxes the system to an equilibrium configuration.  In this case the Zerilli waveform appeared noisy and did not exhibit the quasinormal mode oscillation. The Zerilli perturbation expansion about a background Schwarzschild spacetime was probably affected when scalar radiation moved through the region of the detector. Even when the initial configuration has the detector in an almost Schwarzschild region, the dynamics of the star can push scalar radiation into this region during the evolution. The Newman-Penrose scalar was observed to be more robust in the presence of scalar radiation.

Fig.\ \ref{psi4_old} shows the $\Psi_4$ waveform for this perturbation for a Model 1 
boson star superimposed on the waveform for the same star under the $Y_{20}$ perturbation 
of Fig.\ \ref{zerilli_y20}(b). The R-NR simulation was conducted on a $164^3$ grid with 
resolution $\Delta x=\Delta y=\Delta z=0.35$. Since the $Y_{20}$ perturbation is significantly smaller, it gives a much smaller signal. In order to make a clear comparison the waveform for the 
$Y_{20}$ perturbation has been magnified by a factor of $100$. Although the R-NR 
perturbation is not huge, it is large enough to prevent the star from ringing into 
its quasinormal modes initially. However, it appears to move
into these modes after a few oscillations as can be seen in the figure. The 
oscillation is very clearly damping out on a short timescale as it is characteristic 
for boson stars \cite{futamase}. 

\subsection{Unstable Star Collapse to a Black Hole}
\label{blackhole}
\begin{figure}[ht]
\begin{center}
\leavevmode
\epsfxsize=200pt
\epsfbox{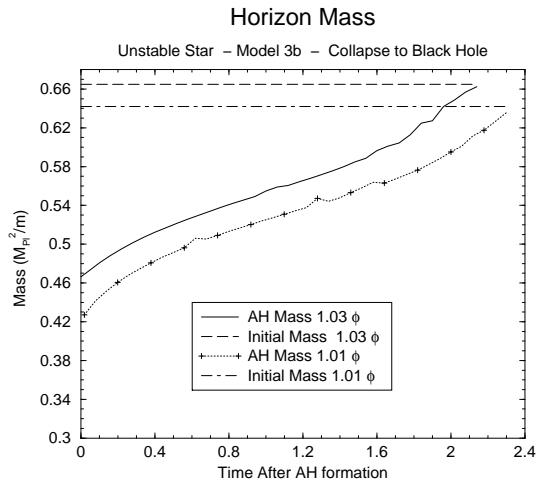}
\caption{The mass of the apparent horizon vs.\ the time elapsed after horizon formation
is displayed for star collapse simulations Run~1 and Run~2 described in the text.  
Run~1 has an amplification factor of $1.01$  and an initial 
mass of $ M = 0.642 M_{\rm{Pl}}^2/m$. 
Run~2 has radial amplification factor of $1.03$ and initial 
mass $ M = 0.665 M_{\rm{Pl}}^2/m $. 
In each case the horizon mass is observed to 
steadily increase towards the total mass of the initial configuration as more 
scalar matter falls through the horizon surface. An asymptotic approach to the value
of the initial mass is expected and is observed at early times, but large errors 
due to grid stretching cause the mass to rise quickly at late time.}
\renewcommand{\arraystretch}{0.75}
\label{collapse}
\end{center}
\end{figure}

\begin{figure}[ht]
\begin{center}
\leavevmode
\epsfxsize=200pt
\epsfbox{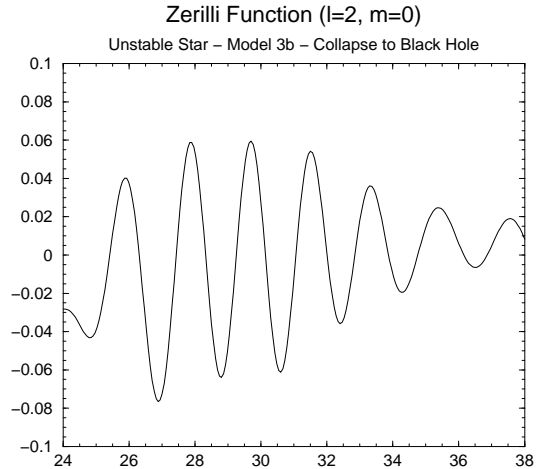}
\caption{The ($\ell=2, m=0$) Zerilli waveform for the 
case of the unstable star of Model 3b with amplification factor of $\mu=1.03$ and
nonradial perturbation specified by $ \epsilon_{20} =  0.128 $ and $ \epsilon_{22} = 0.104 $.}   
\renewcommand{\arraystretch}{0.75}
\label{collwave}
\end{center}
\end{figure}

In this section simulations that follow the collapse of unstable boson stars under nonspherical perturbations to the formation of black hole horizons are presented.
The unstable star designated as Model 3b in Table \ref{models}
is used as the starting configuration. A parabolic weighted nonradial
perturbation centered at $R_p=2.5$ (about half of the $95 \%$ radius of the
unstable star) was applied. This nonradial perturbation was superimposed
on different radial perturbations that accreted additional matter onto the star
by means of a uniform amplification of the field of the form $\mu \, \phi_{\rm{eq}}$ 
(where $\phi_{\rm{eq}}(r)$ is the initial equilibrium  field configuration). 
The addition of scalar matter by a spherical amplification hastens the collapse process. 
In contrast, the star is observed to collapse at approximately the same rate 
with or without the nonradial perturbation.

The time dependence of the mass of the apparent horizon for two collapse 
simulations is displayed in Fig.~\ref{collapse}.  
The nonspherical perturbations for the systems are parametrized by 
$ \epsilon_{20} =  0.128 $ and $ \epsilon_{22} = 0.104 $ (using Eq.\ (\ref{ylm_pert})). 
The first simulation, denoted Run~1, has 
a radial amplification of $\mu = 1.01 $ while the second, denoted Run~2, 
has a radial amplification of $\mu = 1.03 $.  
For these amplifications the mass of the star changes to $ M = 0.642 M_{\rm{Pl}}^2/m $ for the $\mu=1.01$ case
(an increase of about $1.7\%$ from the unperturbed mass)
and $ M = 0.665 M_{\rm{Pl}}^2/m$ for the $\mu=1.03$ case
(an increase of about $5.4\%$ from the unperturbed mass). 
The horizon is observed to form at a time of $t=38.3$ within Run~2. This is earlier than 
the formation time of $t=54.7$ in Run~1, which is appropriate because of the larger initial mass. 
% $ 1.03 \phi_{\rm{eq}} (r) $ radial perturbation.  
It can be seen that the horizon mass is $ 0.463 M_{\rm{Pl}}^{2}/m $ when first detected for the case of 
Run~2.  The evolution is carried forward for an additional time interval $ t=2.1 $ after horizon formation, 
corresponding to approximately $ t = 3M $  of evolution of the black hole.
A smooth asymptotic approach towards the value of the initial mass is observed at early times after
the formation. However, at late times the errors in the simulation are becoming large and this is
manifested in a slight upward turn of the curve along a path that would apparently cross the
axis corresponding to the initial mass.  
These features are also echoed in the results for Run~1,  
where the mass of the apparent horizon takes the value  $ 0.427 M_{\rm{Pl}}^{2}/m $ at formation and increases in an analogous manner. The late time errors are caused by the familiar problem of grid stretching associated with singularity avoiding time slicing \cite{gridstretch0,gridstretch1,gridstretch2}. It should be noted that these simulations did not
make use of advanced shift conditions or excision techniques. Future work will include the employment of 
such methods to extend the duration of the simulations for long times after the black hole formation.

The geometry of the apparent horizons formed in Run~1 and Run~2 is investigated by calculating 
the ratio of the polar to equatorial circumference $ C_r = C_p/C_e $ for two different great circles 
$ \varphi = 0 $ and $ \varphi = \pi / 2 $.  The values for $ C_{r}(\varphi=0) $ and $ C_{r}(\varphi=\pi/2) $ are very close to unity, with values in the range $ C_r = 0.9998 \pm 0.0002 $ on all timesteps for which horizons are detected 
in the simulations. The nearly spherical shape of the horizons
can be understood by considering the Zerilli waveform of Run~2 displayed in Fig.~\ref{collwave}. 
The waveform starts to ring down near $ t = 30$ and damps to a small value by $ t=38 $, 
which precedes the formation time of the apparent horizon within the simulation. 
The nonradial perturbations have decayed away and the system has become nearly 
spherical by the time of horizon formation. 
However, it can be demonstrated that a slightly nonspherical horizon may be 
obtained when the formation occurs before the full relaxation of the nonradial modes.   
Another simulation, Run 3,  takes an initial 
configuration with a very large amplification factor $ \mu = 1.15 $ and the same nonradial coefficients $ ( \epsilon_{20}=0.128, \epsilon_{22}=0.104) $ as Runs 1 and 2. 
The initial mass of this configuration is $ 0.801 M_{\rm{Pl}}^{2}/m $, and the subsequent
evolution results in a horizon formed by $ t = 20.5 $. 
\begin{table}[h]
\begin{center}
\begin{tabular}{|>{\footnotesize}p{1.6in}>{\footnotesize}p{1.6in}|}  \hline
\multicolumn{2}{|c|}{AH circumference ratios} \\ \hline
  $t_{\rm{formation}}$&  $t_{\rm{end}}$  \\ \hline \hline
 $C_{r}(\varphi=0) \, \, \, =1.0017$    &  $C_{r}(\varphi=0)  \, \, = 1.0008$        \\
 $C_{r}(\varphi=\frac{\pi}{2})=1.0009$  & $C_{r}(\varphi=\frac{\pi}{2})  = 1.0004$   \\ \hline
\end{tabular}
\end{center}
\caption{The ratios of the polar to the equatorial circumference for two different great circles ($\varphi=0$ and $\varphi=\pi/2$) are displayed at the horizon formation time and at the end of 
Run 3. The termination time is $ t_{\rm{end}} = t_{\rm{formation}} + 3.2 \approx t_{\rm{formation}} + 4 M $, where the apparent horizon forms at the time $ t_{\rm{formation}}=20.5$.}
\label{c_rs}
\end{table}
%The values of the circumference ratios are 
%\begin{eqnarray}
% t_{\rm{formation}}&:&C_{r}(\varphi=0)  = 1.0017 \quad t_{\rm{end}}:C_{r}(\varphi=0)  = 1.0008 \nonumber \\ 
% t_{\rm{formation}}&:&C_{r}(\varphi=\pi/2)  = 1.0009 \, t_{\rm{end}}:C_{r}(\varphi=\pi/2)  = 1.0004 \nonumber  
%\end{eqnarray}
From the values of the circumference ratios displayed in Table \ref{c_rs} it is seen that the horizons are measurably nonspherical. The circumference ratios are seen to steadily decrease, suggesting the eventual evolution towards spherical symmetry. Furthermore, we see that the perturbation proportional to $ Y_{22} $ has resulted in 
a dependence in the ratio $ C_r $ on the angle $ \varphi $ at which the polar circumference is measured. 
The formation and study of horizons with complex shape from boson star collapse will be investigated 
further by using advanced methods for the evolution of black holes.  

\subsection{Migration from the Unstable Branch}
\label{migration}

A boson star on the unstable branch can migrate to a new configuration of
smaller central field density that lies on the stable branch. This 
behavior is analogous to the behavior of relativistic neutron stars \cite{malcolm}.  In 
both cases there is a large initial expansion of the star followed by oscillations about 
the final stable star configuration it will settle into. In the case of boson stars these 
oscillations damp out slowly, similar to the case of stars with a perfect fluid equations of 
state. The neutron star with perfect fluid equation of state settles into a configuration 
on the stable branch of slightly lower mass than the initial unstable star.  The initial 
zero temperature star is heated due to gravitational binding energy being
converted to internal energy by shock heating. The boson star on the other hand loses 
mass to scalar radiation and settles to a configuration of lower mass on the stable branch.

 The mass of a critical boson star (Model 2, Table \ref{models}) was lowered by about $3.5 \% $ 
through a radial perturbation. This large perturbation reduced the boson 
field everywhere to $98 \% $ of its original value. Next, a nonradial parabolic
weighted perturbation (see Sec.\ \ref{methods}) was superimposed on the radial one.
The weighted perturbation was centered at a radius of $R_p=3.025$, which is about half the
$95\%$ radius of the star. This was a pure $Y_{20}$ perturbation with $\epsilon_{20}=0.016$
in Eq.\ (\ref{ylm_pert}). The data was then passed through the 
IVP and its evolution was tracked using the 3D code.  
 
The strong dynamics within the simulation causes substantial coordinate drifting. A dense 
unstable configuration with a small radius requires good resolution. This star migrates 
to a dilute stable configuration with a larger radius (about twice as large). The different 
scales place demands on both the resolution and gauge choice within the simulation. The 
gauge control was obtained by enforcing maximal slicing through a K-driver restoring term.  
When $K$ is perturbed away from zero, this term drives it back exponentially in time \cite{coodslicebala}. 
 For computational efficiency, the enforcement of maximal slicing
is performed intermittently, with the algebraic 1+log slicing 
condition used on intermediate timesteps. While some gauge
drift persists the coordinates within the simulation are stable.  Future work will involve studying 
better gauge control through an adaptive mesh refinement technique \cite{carpet}.
\begin{figure}[t]
\begin{center}
\leavevmode
\epsfxsize=200pt
\epsfbox{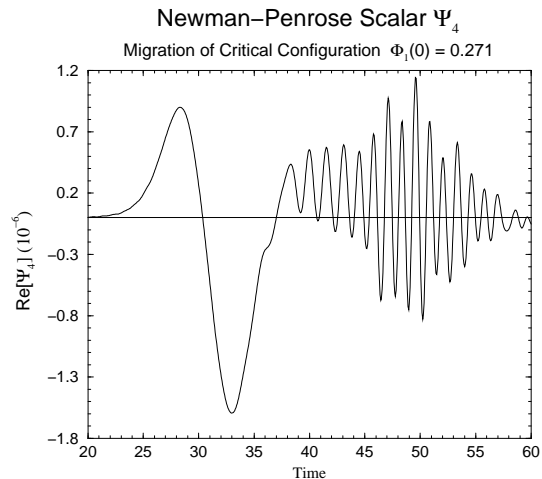}
\caption{The Newman-Penrose waveform $\Psi_4$ is shown for a migrating critical star (Model 2). 
The initial mass of the unperturbed star was $0.633 M_{\rm{Pl}}^2/m$ and after a perturbation 
of the form $\Phi(r) \to 0.98 \Phi(r)$ was lowered to $0.611 M_{\rm{Pl}}^2/m$. On top of this 
spherical perturbation a nonradial parabolic weighted perturbation proportional to $0.016$ $Y_{20}$ 
was applied. This was centered at about half the radius of the critical star. A grid size of 
$164^3$ with resolution $\Delta x = \Delta y = \Delta z = 0.2$ was used. This waveform is significantly different from the 
nonradial quasinormal modes of a critical star. This is expected since the large initial 
perturbation moved the star significantly away from the original critical configuration. 
Subsequently it migrated to the stable branch, moving it further away from its initial 
configuration. }
\renewcommand{\arraystretch}{0.75}
\label{migr_wavef}
\end{center}
\end{figure}

Fig.\ \ref{migr_wavef} shows $ \Psi_4 $ at a detector at $r=30$. The star is dynamic 
and moves away from the initial state even in the short damping timescale of 
the gravitational wave signal. Hence, the gravitational wave does not carry the signature of the original 
configuration. The migration process to a stable star is an extremely lengthy 
process, and the waveform shown is emitted during the very early stage of this 
process. Thus, the waveform does not show the quasinormal modes
of the eventual final state stable star or the initial critical configuration, but 
rather a complex combination of the modes of a mixture of states that the 
system goes through during the migration. A good modefit 
of this data with the quasinormal modes of any single star configuration 
is therefore not possible. The lengthily and complicated precursor (both low 
and high frequency parts) is indicative of the complex dynamics of the strong 
perturbation and migration process. A Fourier analysis of this waveform in the 
ring down region results in a dominant frequency of  $4.3 \, m/M_{\rm{Pl}}^2$. 
This frequency is reflective of the dynamics of the star under a strong 
perturbation, which does not result in excitation of quasinormal modes. It is much higher 
than any of the quasinormal modes of a critical star (as determined by Ref.\ \cite{futamase}).  

Since the detectors have to be far away for $ \Psi_4 $ to represent the gravitational 
signature of the star, a large grid had to be used to obtain the waveform. A very long 
run with such a grid would be computationally expensive. However, due to the strong 
damping of these waves, one can see the full gravitational ringing of the star on a 
short time scale ($t=60$ for this simulation). On the other hand, if one wants to see the star continue its migration for a few radial oscillations, a longer 
run needs to be conducted. The period of one radial oscillation for this configuration 
is a lot larger than the damping time of the gravitational wave (almost four times as large).  A two pronged
approach was used. A shorter run with a large grid ($164^3$ of resolution $\Delta x=\Delta y=\Delta z=0.2$) 
was performed to obtain the gravitational wave. Subsequently, a smaller grid ($96^3$ of resolution $\Delta x=\Delta y=\Delta z=0.2$)  
was used to handle the migration. This was compared to a spherical perturbation run with the same grid and resolution. 

The simulation proceeds accurately for several 
hundred oscillations of the underlying scalar field, with the L2-norm 
of the Hamiltonian constraint on the grid persisting at a value of 
order  $10^{-4}$ at late times. At the time at which the simulation 
is stopped, it has not reach an error prone state at which it must be 
terminated, but rather we simply exhaust our available computer 
time. By all available indicators, the simulation would continue 
further at this level of accuracy. The level of the errors 
in our migration study  compares very favorably 
to other published work. 3D numerical simulations by  
Font {\it et  al.} \cite{malcolm} follow long-term evolutions of relativistic neutron stars 
and show errors that grow to a level of $10^{-2}$ by the end of the simulation. 
The constraint values we report persist at least an order of magnitude below 
this level for hundreds of oscillations of the underlying scalar field, the relevant 
time scale for our problem.
\begin{figure}
\begin{center}
\leavevmode
\epsfxsize=200pt
\epsfbox{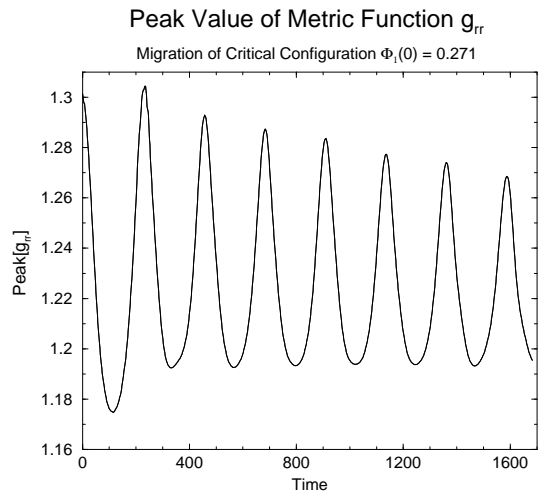}
\caption{The maximum value of metric function $g_{rr}$ is plotted against time for 
the migration of a critical star to the stable branch. The initial mass of the 
star was $0.633 M_{\rm{Pl}}^2/m$ and after a perturbation of the 
form $\Phi(r) \to 0.98 \Phi(r)$ was lowered to $0.611 M_{\rm{Pl}}^2/m$.
The nonradial oscillations have damped out early on (approximately by time $t=60$)
and the remaining spherical oscillations proceed through seven periods within this simulation. 
The frequency is very steady by the end with the last four peaks at $t=909, 1135, 1361,$ and $1587$, all separated equally by $\Delta t = 226$. A $96^3$ grid of resolution $\Delta x = \Delta y = \Delta z =0.2$ was used. }
\renewcommand{\arraystretch}{0.75}
\label{grr_migration}
\end{center}
\end{figure}

Fig.\ \ref{grr_migration} shows the maximal radial metric $g_{rr}$ as a function of 
time for a migrating (Model 2) critical star. The initial drop in the metric indicates 
a migration to the stable branch. The 
star then oscillates about the new stable configuration that it will settle into 
with the radial oscillations damping out slowly. The star settles into a fairly 
constant oscillation frequency.  The last four peaks are at $t=909, 1135, 1361,$ and $1587$, 
all separated by $\Delta t = 226$. The metric in Fig.\ \ref{grr_migration} also exhibits 
a shift upwards. We believe this a nonphysical effect, likely a result of gauge drift 
that occurs in response to the very large initial perturbation. One can roughly estimate 
the final configuration to which the star will settle. The final star configuration of mass 
$0.59 M_{\rm{Pl}}^2/m$ corresponds to a maximum metric of 
$g_{rr \, \rm{max}} = 1.18$. 

Since the star expands to a configuration of larger radius the grid size 
had to accommodate this expansion. This was at the expense of the grid 
resolution of the initial denser, more compact configuration. Similar 
problems were seen for neutron star simulations. In the neutron star 
case a $96^3$ grid was used to study the migration and the low 
grid resolution of the initial configuration caused a nonnegligible deviation 
of the average central rest mass density from the expected value \cite{malcolm}. 
We plan to investigate the use of fixed mesh 
refinement \cite{carpet} to accommodate the multiple length scales 
that occur within such migration studies.

In the nonradial perturbation case the high frequency  oscillations,         
associated with the nonradial modes, of the peak of the metric function 
$g_{rr}$ are superimposed on the low frequency
radial oscillation. This superposition is not clear in the figure because the 
small amplitude of nonspherical oscillation is suppressed by the larger spherical 
oscillation amplitude. The spherical oscillation amplitude is of order $0.1$. This 
is much larger then the corresponding radial oscillation amplitude due to discretization 
errors (of order $0.003$) from Fig.\ \ref{gxx}(a) for the same timescale. While the nonspherical oscillation can not readily be seen, the metric exhibits the same superposition of the radial and nonradial frequencies discussed in Fig.\ \ref{gxx}(a).

\section{Conclusion}
This is the first paper to present the long time evolutions of boson stars under 
nonradial perturbations in full 3D general relativity. The existence of rapidly 
damping quasinormal modes as discussed by Ref.\ \cite{futamase} is confirmed. For the 
first time gravitational waveforms (both Zerilli and Newman-Penrose scalar $\Psi_4$) 
have been fully extracted and presented for boson stars in 3D. 
%Under small spherical perturbations a star expands and contracts at its radial quasinormal mode frequency and loses mass at each expansion through scalar radiation. In conjunction with the expansion and contraction of the star the metric functions undergo low frequency oscillations that damp off on a large timescale \cite{edw1}. 
We find that when a small nonradial perturbation is applied to a boson star, the star loses mass through gravitational radiation and has quasinormal modes 
frequencies that are much higher than its radial quasinormal mode frequency. The 
metric acquires this frequency signature. If the small nonradial perturbation 
in the scalar field is doubled the metric perturbation also doubles. For perturbations that 
have radial and nonradial components the metric exhibits both frequencies superimposed 
on one another. The gravitational wave signal is observed to damp out quickly and the star becomes spherical on a short timescale.

In this paper, the gravitational waveforms have been presented for stable, critical 
and unstable boson star configurations. Our results were compared to linear 
perturbation results of Ref.\ \cite{futamase}. However, the comparison could only 
be approximate because their WKB formulation leads to inaccuracy for the calculation 
of the lowest modes. Furthermore, the calculation of the quasinormal frequencies in
 Ref.\ \cite{futamase} is slightly dependent on the choice of the surface of the star. 
 In principle, a boson star is of infinite extent with exponential damping of the scalar 
 field within a short radius. This exponential damping allows one to consider a large 
 part of the star to be virtually a vacuum and arbitrarily choose a surface somewhere in this region 
for calculational (perturbation theory) or numerical (finite grid) purposes. 
Different choices of surface can lead to slightly different results for the modes (changes them 
within a few percent). 
%Our numerical code allows virtually any type of perturbation.

We have considered perturbations that are purely nonradial and are proportional 
to the spherical harmonics. The Zerilli gravitational waveforms of the $\ell=2, m=0$ 
mode for small perturbations proportional to $Y_{20}$ have the same 
frequencies as waveforms generated from perturbations composed of a linear 
combination of $Y_{20}$ and $Y_{22}$  for a given star configuration. This 
shows that a given star has a specific quasinormal mode signature.

It is more challenging to obtain gravitational waveforms when matter is 
present in the spacetime. Even if the system has an approximate Schwarzschild 
exterior region the dynamics of the simulation might carry matter there. The 
presence of matter can affect the Zerilli waveform which requires a region 
that is close to Schwarzschild. This effect was tested explicitly by using 
different types of perturbations. The first was purely nonradial, while the second 
was more general and had radial and nonradial components. The former did not 
result in scalar radiation and the Zerilli waveform was observed to exhibit the high
 frequency quasinormal mode oscillation, while the waveform in the latter case 
 was noisy and seemed affected by the presence of scalar radiation. 
 The Newman-Penrose waveforms behaved better in all cases.
%The first was purely nonradial, while the second was more general and had radial and nonradial components. The former 
%did not result in scalar radiation and gave excellent Zerilli waveforms, while 
%the waveform in the latter case was affected by the presence of scalar radiation. 
%The Newman-Penrose waveforms behaved better in all cases.

Nonspherical apparent horizons were observed under nonradial perturbations when 
the horizon formed before the full emission of the gravitational waveform. Radial 
perturbations that amplify the mass of the star are seen to accelerate the collapse. 
However, the nonradial perturbations do not affect the collapse time. The geometry 
of the horizon is seen to become more spherical as the star evolves. Even for large 
nonradial perturbations the degree of asymmetry is quite small.

Under large radial perturbations that mimic the removal of scalar field, an unstable 
branch star can migrate to the stable branch \cite{edw1}. A large radial perturbation 
of an unstable star was performed for the first time in 3D and several oscillations 
of the metric were observed with the star settling to a constant oscillation 
frequency. This is a dynamic problem that involves multiple scales. 
Gauge control was obtained using a stable implementation of maximal slicing. 
Improvements are under study. A nonradial perturbation superimposed on the radial 
perturbation resulted in the emission of gravitational waves during the migration. 
The migration of an unstable branch star to the stable branch under a 
general perturbation (with both radial and nonradial components) can be significant 
in the formation of boson stars \cite{edw1}.
\section*{Acknowledgments}
We thank Jason Ventrella and Manuel Tiglio for the careful review of our manuscript. 
Our code is fully based on the Cactus Computational Toolkit.
We gratefully acknowledge the Cactus team for their help and extensive support, particularly Thomas Radke. We want to especially thank Gabrielle Allen for her assistance and overall guidance. We would like to recognize Erik Schnetter for developing the Cactus TAT interface 
to the PETSc elliptic solver that is used by our IVP solver. We thank Jian Tao, Malcolm Tobias, and Wai-Mo Suen for useful conversations and access to Washington University computer resources. J.B. would like to especially thank Matt Visser for helpful discussions and encouragement for this project. R.B. is grateful to Doina, Cornel, Ileana and Ruxandra Costescu for their hospitality and support during the length of this project. The large-scale computations were performed on the Platinum and Tungsten clusters at NCSA under the NSF NRAC grant MCA02N014 and on the SuperMike cluster at LSU. We thank LSU CCT for their well-organized visitor program. We also acknowledge generous support from Microsoft. F. S. G. and E. S. acknowledge partial support from the bilateral project DFG-CONACYT 444 MEX-13/17/0-1. F.S.G. is partly 
granted by CIC-UMSNH-4.9 and PROMEP-UMICH-PTC-121.

 \end{document}